 \documentclass{emulateapj}
\usepackage{mathtools}
\usepackage{natbib}
\usepackage{CJKutf8}
\usepackage{graphicx,amsmath,amsthm,ulem,color}
\usepackage{color}
\usepackage{amssymb}
\makeatletter

\newcommand{\red}{\textcolor{black}}

\begin{document}

\title{Habitability from Tidally-Induced Tectonics}

\author{ Diana Valencia$^{1,2}$,  Vivian Yun Yan Tan$^3$ and Zachary Zajac${^4}$}
\affil{$^1$  Department of Physical $\&$ Environmental Sciences,
  University of Toronto, Toronto, Canada \\
 $^2$ Department of Astronomy \& Astrophysics, University of
  Toronto, Toronto, Canada\\
 $^3$ Department of Physics and Astronomy, York University \\ 
 $^4$ Department of Medical Biophysics, University of Toronto}

\begin{abstract}
The stability of Earth's climate on geological timescales is enabled
by the carbon-silicate cycle that acts as a negative feedback mechanism
stabilizing surface temperatures via the intake and outgas of atmospheric
carbon. On Earth, this thermostat is enabled by plate tectonics that
sequesters outgassed CO$_{2}$ back into the mantle via weathering
and subduction at convergent margins. Here we propose a separate tectonic
mechanism --- vertical recycling --- that can serve as the vehicle for CO$_{2}$
outgassing and sequestration over long timescales. The mechanism requires
continuous tidal heating, which makes it particularly relevant to
planets in the habitable zone of M stars. Dynamical models of this
vertical recycling scenario and stability analysis show that temperate climates stable over
Gy timescales are realized for a variety of initial conditions, even
as the M star dims over time. The magnitude of equilibrium surface temperatures
depends on the interplay of sea weathering and outgassing, which in turn depends on planetary carbon content, so that
planets with lower carbon budgets are favoured for temperate conditions. 
Habitability of planets such as found
in the Trappist-1 may be rooted in tidally-driven tectonics.
\end{abstract}

\section{Introduction}

In the search for habitable planets, we rely on our knowledge of the
Earth to guide us. We understand that it is not enough for a planet
to be located in the habitable zone for it to be habitable. It is
equally important that its atmospheric response to insolation allows
for liquid water on its surface, and this depends on the amount and
type of greenhouse gases present. 

The long term stability on Earth has been attributed to the carbon-silicate
cycle, that maintains atmospheric carbon dioxide levels at values
that allow for surface liquid water over million-year timescales,
while exchanging carbon between the different major reservoirs (atmosphere
and ocean, continental and oceanic crusts, mantle). The main reason
that this cycle brings climate stability, is that weathering from the
atmosphere depends on atmospheric temperature. When temperatures rise, weathering
rates increase, drawing down CO$_{2}$ from the atmosphere into the
rocks, thus reducing the greenhouse effect and restoring temperature
levels. Conversely, when the temperature is cold, weathering is sluggish
or non-existent (if the planet has gone into a snowball state), allowing
for volcanism to increase levels of CO$_{2}$ in the atmosphere. 

Evidence in the geological record suggests that the Earth's climate
has been temperate over most of the last 3-4 billions years, despite
the fact that the Sun has been brightening over time \citep{Sagan:1972}.
\citet{Owen:1979,Kasting:1993} proposed that higher levels of CO$_{2}$
in the past, possibly due to the carbon-silicate cycle, could offset
the reduced insolation level.  While this is the leading theory, studies against
the CO$_2$ being able to resolve the faint young sun paradox include limits to the amount
of atmospheric CO$_2$ in the past derived from siderate palaeosols data \citep{Rye:1995} as well
as inferences from modelling vigorously convecting mantles and reactable ejecta in the
early earth that would draw down atmospheric CO$_2$ to
too low a value \citep{Sleep:2001}.  While the need for other greenhouses such as NH$_3$
 (suggested by \citet{Sagan:NH3}) might be the answer to these caveats, 
the carbon-silicate cycle on Earth, with the ability
to regulate atmospheric CO$_{2}$, has at least
to some extent contributed to the long term
climate stability of our planet.

It is also true that in the case of the Earth, this cycle is enabled
by the fact that plate tectonics connects the different reservoirs.
Carbon is drawn from the atmosphere into the rocks via rock weathering
on the continents and sea weathering on the ocean crust. Through rivers
and streams, continental rocks get deposited into the oceanic crust,
which gets subducted into the mantle at convergent margins, while
continental crust is scrapped and carried down by the subducting plate.
Through volcanism, carbon is outgassed from the mantle into the ocean
at mid-ocean ridges, and directly into the atmosphere at continental
arcs and ocean islands. Thus, subduction, which is a central component
to plate tectonics, closes the carbon-silicate cycle on Earth. 

In addition, plate tectonics is also important to the carbon-silicate
(C-Si) cycle because it assists the weathering process by constantly
exposing fresh rock subsequently available for carbon sequestration, either
by continually producing ocean crust at the mid-ocean ridges and ocean
islands, or by enabling erosion on the continents through persistent
topographical changes derived from mountain building and orogeny processes
\citep{Geodynamics:Book}. Given these reasons, plate tectonics has
been tied to climate stability and hence, habitability on Earth \citep{Walker:1981, 
KumpArthur, Gaillardet:weathering, West:2005, West:2012, MaherChamberlain}. 

However, it is debated whether or not plate tectonics can happen in
exo-Earths. With some suggesting it is possible \citep{Valencia:PT,VanHeck:PT,Foley:PT,Korenaga:PT,Tackley:PT},
while others consider it to be unlikely \citep{ONeill:PT,Stamen:PT,Noack:PT}.
In this study, we explore a different type of tectonism, driven by
tidal heating, that may serve in an analogous way to plate tectonics
on Earth in assisting a carbon-silicate cycle. Thus, increasing the
chances of finding planets that are habitable. 

Inspired by the efforts on finding planets in the habitable zone around
M stars, where tidal heating can be important, we envision a tectonic
scenario where volcanism is driven by tidal dissipation within the
mantle, in a similar fashion to what has been proposed for Io \citep{OD:Io}. 

This mechanism would be highly relevant for planets that have non-zero
eccentricities orbiting in the habitable zones of M stars, such as
three of the seven planets of the Trappist-one system \citep{Trappist-1}.
This recently discovered system of seven highly packed planets near
resonances, includes three $d,e,f$ in the habitable zone. Given the
planets' gravitational perturbations with its neighbours, we expect
small nonzero eccentricity values \citep{Hansen:2015,Hansen:2017,Tamayo:2017}
making these three planets highly suitable candidates to exhibit tidally-driven
tectonics, and perhaps a built-in climate stability thermostat analogous
to Earth. 

In this manuscript we investigate how this newly proposed mechanism
may enable climate stability for planets that are tidally heated and
are found in the habitable zone. It is organized as followed: in section 2 we present
the tidally driven tectonism we envision and how it can assist climate
stability, as well as the governing equations; in section 3 we discuss
the results; in section 4 we discuss our assumptions and implications,
and present a summary of our findings in section 5.

\section{Model}

\subsection{Tidally Driven C-Si Cycle Scenario}

To come up with an alternative system, we need to break down the key
elements of the carbon-silicate cycle on Earth that provide our planet
with a viable thermostat for climate stability. (1) There needs to be
a feedback mechanism that draws out CO$_{2}$ from the atmosphere
that varies with CO$_{2}$ concentrations \citep{Walker:1981}, and
deposits it in a different reservoir. On the case of the Earth this
is both the rock and sea weathering processes that depend on CO$_{2}$concentrations
directly, and very importantly, indirectly via the atmospheric temperature,
and store carbon in the rocks and ocean crust. (2) This mechanism
has to supply fresh rock for weathering at a rate large enough rate as to
not produce a bottleneck in the system. On Earth this rock exposure happens continuously thanks to 
persistent erosion and mid ocean ridge production . (3) And lastly, the reservoir has to have a way of injecting
CO$_{2}$ back into the atmosphere when atmospheric levels decrease.
On Earth this happens because volcanism is a continuous source of CO$_{2}$
from the mantle, and the mantle in turn is continuously replenished
with subducted carbonate rocks from the ocean and continental crust. 

\begin{figure}
\includegraphics[width=0.45\textwidth,trim=0 0 0 0, clip]{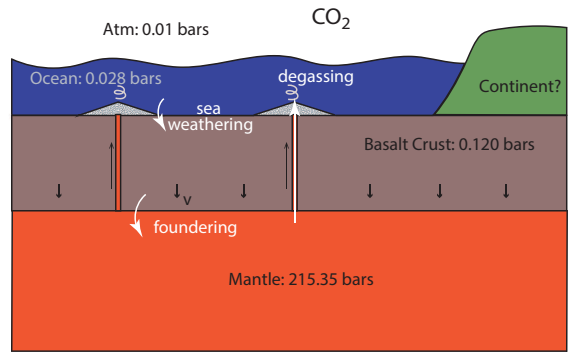}

\caption{Tidally-driven tectonism and pCO$_{2}$ long-term feedbacks. Continuous
volcanism from a partially molten mantle degasses CO$_{2}$ in the
ocean/atmosphere. The basaltic layer that forms from this volcanism
can sequester C from the ocean/atmosphere reservoir via sea weathering.
The basaltic crust grows at the top with resurfaced basalt, and founders
at the bottom moving C into the mantle. There is no need for continents,
and continental weathering, for this mechanism to enable the C-Si
cycle on tidally heated planets. \label{fig:cartoon} }

\end{figure}

An analogous system that accomplishes all four elements is inspired
by the tidally driven tectonism suggested on Io \citep{OD:Io}, or pipe heating and
depicted in Fig. \ref{fig:cartoon}. 

Planets that are tidally heated can dissipate heat in their interior
exhibiting partially molten mantles, that get rid of their heat by
pushing melt through plumes to the surface. This melt continuously
resurfaces in the form of basalt, and forms a layer that accumulates
over time. This freshly advected rock can react with atmospheric or
ocean CO$_{2}$ in a similar way to that on Earth and sequester C
via a weathering reaction. With time, new basaltic crust gets deposited
on top, so that the old basalt and carbonate rock get buried and move
deeper within the planet. At some point, depending on the resurfacing
rate, these rocks that were once at the surface will be buried and
delaminated into the mantle, carrying down carbon with them and closing
the cycle. Thus, recycling occurs in a vertical fashion instead of
the horizontal character of plate tectonics on Earth.

We shall examine in more detail each of the components of this proposed
system before presenting the climate model we built for these planets.

\subsubsection{Tidal Heating}

For a planet with a continuous and substantial source of tidal heating,
the planet may have a partially molten interior \footnote[1]{We note that there is no need 
to invoke a global molten layer, but the presence of partial melt is enough.}. Just like the volcanism
on Io, this melt may reach the surface through plumes advecting heat
out to the surface. It is thus important to calculate the amount of
heating a planet can experience. Based on the theory of tides \citep{MurrayDermott}
the tidal heating available to a planet depends on the eccentricity
of the system $e$, the semi-major axis $a$, the mass of the star
$M_{\star}$, the mean motion $n$, the radius of the planet $R_{p}$,
the specific dissipation parameter $Q$, the ratio of elastic to gravitational
forces $\bar{\mu}$ $\left(\approx(10^{4}\mathrm{km}/R_{p})^{2}\right)$
and the gravitational constant $G$:

\begin{equation}
\frac{dE}{dt}=\frac{63}{4}\frac{e^{2}n}{\bar{\mu}Q}\left(\frac{R_{p}}{a}\right)^{5}\frac{GM_{\star}^{2}}{a}.\label{eq:tidal}
\end{equation}

The least well constrained parameter is $Q$ although typical values
of $\sim100$ are commonly used for icy/rocky planets. In reality
the dissipation parameter is a function of the planet's structure
and should vary radially instead of being a single value. However,
as a first step, we use the same approach as in most studies and use
a constant value. Equation \ref{eq:tidal} can also be written as 

\begin{equation}
\begin{aligned}
\frac{dE}{dt}\approx & 10\times\left(\frac{e}{0.01}\right)^{2}\left(\frac{100}{Q}\right) \\
                                &  \left(\frac{R_{p}}{R_{E}}\right)^{7}\left(\frac{M_{\star}}{M_{\odot}}\right)^{5/2}\left(\frac{0.1\mathrm{AU}}{a}\right)^{15/2}[\mathrm{TW}] \label{dEdt}.
\end{aligned}
\end{equation}

Therefore, an Earth-sized planet orbiting a Sun-like star at $0.1$
AU and low eccentricity of $0.01$, would dissipate 10 TW or $\sim 0.02$
W/m$^{2}$, a tidal heatflux 100 times less than Io's flux \citep{q_Io:Veeder,Io:Spencer2000}.
For comparison, Earth's flux is estimated at $0.09$ W/m$^{2}$ \citep{q_earth}.
To reproduce the measured heat flux for Io at its location with respect to Jupiter with Eq. \ref{eq:tidal}, the dissipation factor
would have to be chosen to be $Q_{\mathrm{Io}}=150.$ 

Given that we are interested in planets in the habitable zone that
 can experience tidal heating, we need to consider planets around M
or even Brown Dwarfs. Planets around G type stars that are in the
habitable zone, are too far away to experience any tidal heating from
the star, whereas M stars have their habitable zone close enough that
tidal dissipation can matter. In fact, tidal dissipation in planets
around M dwarfs has been previously studied in the context of orbital
and thermal evolution \citep{Driscoll:2015}. In fact, \citet{Barnes:TidalLimits}
claimed that there is a limit to how much tidal heating a habitable
planet may experience based on the assumption that too much volcanism
and high resurfacing rates would be inhospitable. In contrast, we
propose that at these conditions, a new mechanism for climate regulation
may kick-in rendering the planet habitable. 

For a star with $M_{\star}=0.08\,M_{\oplus},$ like Trappist-1, an
Earth-sized planet orbiting at $0.02$ AU and $0.01$ eccentricity
would experience $3000$ TW or 6 W/m$^{2}$ of tidal energy. For an
assumed eccentricity of 0.01 planets \emph{d} and \emph{e} would have
a tidal flux of 0.96 and 0.30 W/m$^{2}$, respectively. These numbers quickly grow
with eccentricity, so that planets around M stars can have large tidal
heating fluxes. The important aspect for the model we propose is to
determine if there is enough energy to yield a partially molten mantle.
One could use Io for a simple comparison to infer whether or not planets
like Trappist-1 $d,e$ and $f$ could have a partially molten mantle by
comparing the amount of tidal energy available to the tidal energy
carried to the surface. 

Tidal production according to Eq. \ref{eq:tidal} increases as $R_{p}^{5}$,
while the heat flow carried from the mantle to the atmosphere increases
as $R_{p}^{2}$, so that at zero order larger planets should be more likely to have
an interior melt region. Even at very low eccentricity values of $0.01$,
Trappist 1d would experience similar tidal heat fluxes to that of
Io. Thus, it is possible that the Trappist-1 planets in the habitable
zone have persistent melt in their interior. 

This melt would be advected in the form of volcanic conduits cooling
at the surface, carrying with it volatiles and degassing them into
the ocean/atmosphere (see Fig. \ref{fig:cartoon}). This continuous
basalt extraction would form a layer of a certain thickness, 
limited by a partially molten mantle at the bottom. To estimate
both the thickness of the basalt layer and the resurfacing velocity
that are used in the climate model we use the model by \citet{OD:Io}.

\begin{figure}

\includegraphics[width=0.45\textwidth,trim=0 0 0 0, clip]{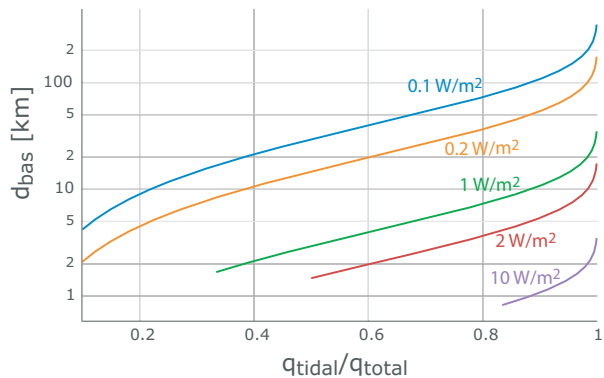}

\caption{Basaltic layer thickness as a function of ratio of tidal heat
  flux to total heat flux for different values of tidal heat flux. The corresponding
  resurfacing velocities are 0.6, 1.2, 6.0, 12, and 60 cm/y for tidal
  fluxes shown in figure from 0.1 to 10 W/m$^2$, respectively. The
  contribution of tidal heat flux to total heat flux $a$ depends on
  the thermal evolution of the planet. \label{fig:v_dbas} }

\end{figure}

The resurfacing velocity, taken to be the same as the subsidence velocity,
is 

\begin{equation}
v=\frac{q_{t}}{\rho\left(L+C_{p}\Delta T\right)} \label{eq:v_res},
\end{equation}

where $q_{t}$ is the tidal heat flux (tidal energy per unit surface
area), $\rho$ is the density of the mantle, $L$ the latent heat,
$C_{p}$ the heat capacity, and $\Delta T=T_{m}-T_{s}$ the difference
between the melting and surface temperature. The thickness of the basalt
layer is 
\begin{equation}
d_{\mathrm{bas}}=\frac{\kappa}{v}\log\left(1+\frac{kv}{\kappa}\frac{\Delta T}{q_{t}\left(\frac{1}{a}-1\right)}\right) \label{d_bas},
\end{equation}
where $\kappa$ is the thermal conductivity, $k$ is the thermal diffusivity,
and $a$ is the ratio of tidal heat flux to total heat flux. While
the resurfacing rate depends only on the tidal heat flux, the thickness
of the basalt layer is also determined by the contribution of tidal energy
flux to total flux, $a$, which is unknown. \red{To illustrate this point, we
have calculated the resurfacing velocity and basalt layer thickness
for different values of $a$. Figure \ref{fig:v_dbas} shows what this
thickness would be as a function of contribution of tidal
heat to the total heat flux. As tidal heating becomes the dominant source
(i.e. $a$ approaches $1$) the thickness becomes infinitely large (as
shown by \citet{OD:Io}) We have also set a minimum contribution by
considering steady state where the total heat flux is from tidal
heating and radioactive decay. By setting a maximum to the radioactive
decay of $2$ W/m$^2$ calculated from radioactive sources on Earth at
3.8 By ago, we obtain a minimum value of $a$. Trappist-1$d$, with a
calculated tidal flux of $\sim 1$ W/m$^2$ according to this simple
model, would have resurfacing velocities of 6 cm/y and a crustal layer
of at least 2 km.}

However, without a detailed calculation
for the thermal history of these planets, which is beyond the scope
of this paper, it is not possible to know
how much of the total flux is coming from tidal heating, so as a simple
first step we take $v$ and $d_{bas}$ to be set independent of each
other. 

\paragraph{Melt} Our model also requires melt to be continuously advected to the surface, 
which depends on how deep the melt is produced, and the chemistry of the melt. 
On Earth, due to plate tectonics, as upper mantle material adiabatically ascends into mid-ocean ridges,
decompression melting takes place near the surface and melt gets carried out as volcanism.  
However, if melt is produced at depth, due to the crossover density of melt and solid at pressures of $\sim 13$ GPa 
or $\sim 350$ km on Earth, melt can stay within the mantle \citep{Ohtani:melt}. In fact, \citet{Noack:melt} has argued on the basis of this
crossover density that stagnant-lid planets with large masses, or large core-mass fractions would not exhibit volcanism -- 
owing to larger surface gravities, the pressure for the crossover density would be reached at shallower depths.
This could pose a problem for our proposed mechanism if tidal dissipation creates melt deeper than the crossover density for the 
specific planet. The pressure for this crossover density depends on the composition of the solid 
(e.g. basalt versus peridotite, or iron content), and the water content of magma \citep{Sakamaki:melt}. The higher the iron content of the solid
or drier the magma, the shallower the crossover density.
\citet{Sakamaki:melt} shows that for Earth, hydrated magma occurring from melting mid-ocean ridge basalt rock has a crossover density
near $410$ km ($\sim 24$ GPa ) if water is present at the 2\% level,
and deeper at the 8\% level. By extrapolating their data and using the
preliminary reference earth model \citep{PREM},   we obtain a crossover density depth for 8\% water in magma of $1400$ km ($\sim 60$ GPa) for Earth. Given as well that tidal strain is largest near the surface, it is reasonable to think that   
tidal heating may produce melt near the surface, and thus our proposed mechanism may work for a subset of planets.

\subsubsection{Sea Weathering}

In broad strokes, the carbon-silicate cycle is the process by which
CO$_{2}$ is drawn from the atmosphere by removing C when interacting
with silicate rocks via water and carbonic acid that releases bicarbonate,
Ca$^{++}$ (and Mg$^{++}$) ions either on the continents or interacting
directly with basalt on the newly formed ocean crust. If on the continents,
these ions eventually find their way to the ocean floor carried down
by rivers and ground water. On the ocean floor they form carbonates
that trap carbon (that was once in the atmosphere) into rocks, that
eventually get subducted into the mantle at convergent margins. The
net reaction is captured in the Urey reaction \citep{Urey:CO2}:

\begin{equation}
\mathrm{CO_{2}(g)+CaSiO_{2}(s)\Longleftrightarrow SiO_{2}(s)+CaCO_{3}(s)} \label{reaction},
\end{equation}

where the right arrow describes how carbonates are formed. Once at
depth these carbonate rocks can react back via metamorphism, releasing
carbon dioxide that eventually makes it back in to the atmosphere
via volcanism (the reverse reaction, left arrow).

On Earth, this cycle is enabled by plate tectonics that continuously
exposes fresh rock on the ocean floor, as well as on the continents
driven by erosion via orogeny and topography build-up. Rock weathering
has been well studied \citep{Walker:1981,Berner:1997,Berner:Book}
while sea weathering has gained more traction in recent years \citep{CooganGillis:2013,CooganDosso:2015}.
The carbonate reactions of sea weathering may be controlled in a different
manner to continental weathering with secondary carbonates and alkalinity
playing an important role \citep{CooganDosso:2015}. It is yet unknown
the contribution of each weathering type to the total weathering process
on Earth, or how it might have changed throughout Earth's history
\citep{Mills:2014}. Although, recent estimates by \citet{KT_Catling:2017}
suggest that continental weathering has been dominant in the last 100 My. 

For the planets proposed here, we envision that tidally-driven tectonism
can continuously expose fresh rock under an ocean, but that any continental
shelves, if present, may not exhibit rock weathering owing to a sustained
lack of erosion. In this case, the only mechanism for drawing down CO$_{2}$
in these planets would be sea weathering. 

While some authors have considered sea weathering as dependent on
CO$_{2}$ concentrations alone \citep{Sleep:2001,Foley:Area}, laboratory
experiments \citep{BradyGislason:1997} and isotopic constraints from
oceanic carbonates \citep{CooganDosso:2015} suggest a temperature
dependence in the form of an arrhenious law. This dependence on the
deep ocean temperature, where Ca leaching, and carbonate production
are taking place depends in turn, on the atmospheric temperature,
and hence the atmospheric CO$_{2}$ concentrations, because atmospheric
temperature determines how much cold surface water enters the thermohaline
circulation system \citep{BradyGislason:1997}. This sea weathering
dependence on atmospheric temperature and CO$_{2}$ is crucial in
allowing for a climate stability feedback. We adopt the same equation
as \citet{Mills:2014} to describe sea weathering as a function of
atmospheric CO$_{2}$ levels and atmospheric temperature $T$ 

\begin{equation}
W_{sea}= \omega W_{sea}^{E}\left(\frac{p\mathrm{CO_{2}}^{atm}}{p_{E}}\right)^{\alpha}\exp\left(\frac{E}{R}\left(\frac{1}{T_{E}}-\frac{1}{T}\right)\right),\label{eq:Wsea}
\end{equation}

where $R$ is the gas constant, $E$ is the activation energy, $W_{sea}^{E}$
is the sea weathering rate baseline estimate (in bar/My or mol/My)
for \red{ a reference state that we set at first to be} the equilibrium atmospheric partial pressure of carbon dioxide
$p_{E}$, and atmospheric equilibrium temperature $T_{E}$\red{, and $\omega$
is a function that lumps all other quantities that sea weathering
might depend on}. The feedback
mechanism comes from the dependency on atmospheric temperature. Any
deviations from the equilibrium temperature would drive much higher
or lower levels of sea weathering depending on whether the planet
is hot with high levels of atmospheric CO$_{2}$ or cold, respectively. 

\red{In addition, sea weathering may include a dependency (through $\omega$)} on the velocity at which fresh rock is exposed,
which in the case of the Earth is the ridge spreading velocity, which has
has changed over time. \red{For Earth this term is often written as
$\omega = \left( \frac{v}{v_\oplus} \right)^\beta $, where $v_\oplus$ is the present day
velocity. While many authors \citep{Sleep:2001, Mills:2014, Foley:Area} use $\beta =1$, \citet{KT_Catling:2017}
propose $\beta \sim 0.5 $. In either case,  this contribution is of order unity.  Other
factors may also be affecting weathering rate (e.g. alkalinity, grain
size, etc), and thus for simplicity
we take $\omega \sim 1$; or conversely our results for sea
weathering can be taken as scaling with $\omega$.  }

Two terms in Eq. \ref{eq:Wsea} are poorly known, the direct dependence
on atmospheric carbon dioxide pressure $\alpha$ and the activation
energy $E.$ \citet{BradyGislason:1997} proposed $\alpha=0.23$ and
$E=41$ kJ/mol, while \citet{CooganDosso:2015} proposed $E=92$ kJ/mol,
and \citet{KT_Catling:2017} considered a range of $E=40 - 110$ kJ/mol. We vary
both $\alpha$  and $E$, \red{as well as perform a stability analysis} to see
the effect on the climate stabilizing feedback we propose here.

\paragraph{Alkalinity}

\citet{CooganDosso:2015} have argued that carbonate mineral precipitation
is largely controlled by alkalinity, and the fact that rock dissolution
involved in sea weathering increases alkalinity, the effect is to
efficiently drive carbon sequestration. We note that Eq. \ref{eq:Wsea} does
not have an explicit dependence on alkalinity, but an implicit one,
as it depends on global temperatures that influence deep water ocean
temperature, which in turn controls the dissolution rates of mafic
and ultramafic rocks \citep{KT_Catling:2017}.

\subsection{Governing Equations}

\subsubsection{Carbon cycle}

To build our carbon cycle model we divide our planet into distinct
reservoirs: the mantle, basalt layer, and treat the ocean and atmosphere
together (similar to \citep{Sleep:2001,Foley:Area}). See Fig. \ref{fig:cartoon}
for a cartoon representation of the tectonic process we propose in
this study.

Magmatic volcanism originates in the molten mantle layer and carries
melt within pipes that outgas CO$_{2}$ into the ocean and atmosphere
reservoirs at a rate proportional to the resurfacing rate $v$. The
flux of CO$_{2}$ degassed into the atmosphere-ocean reservoir is 

\begin{equation}
D=p\mathrm{CO_{2}}^{man}\left(\frac{fvA_{p}}{V_{man}}\right) \label{degas}, 
\end{equation}

where $p\mathrm{CO_{2}}^{man}$ is the partial pressure of carbon
dioxide in the mantle, $V_{man}$ is the volume of the mantle, $A_{p}$
is the area of the planet, and $f$ is the fraction of CO$_{2}$ within
the melt that gets outgassed. On the other hand, the sink for the
atmosphere-ocean reservoir is the flux of CO$_{2}$ that is weathered
at the bottom of the ocean. Namely, the sea weathering rate described
in Eq. \ref{eq:Wsea}. By drawing out CO$_{2}$ from the ocean, sea
weathering effectively draws out CO$_{2}$ from the atmosphere given
that the partitioning between the two reservoirs is set by how soluble
CO$_{2}$ is. Following Foley et al 2015, we use Henry's law for solubility
to determine how much CO$_{2}$ is in the atmosphere ($p\mathrm{CO_{2}}^{atm}$)
versus the ocean ($p\mathrm{CO_{2}}^{oc}$),

\begin{equation}
\begin{aligned}
p\mathrm{CO_{2}}^{atm} = & \, p\mathrm{CO_{2}}^{atm+oc}-p\mathrm{CO_{2}}^{oc} \\
                                        = & \, \frac{K_{H}}{\mu_{\mathrm{CO_{2}}}}\frac{p\mathrm{CO_{2}}^{oc}}{\left(\frac{p\mathrm{H_{2}O}}{\mu_{\mathrm{H_{2}O}}}+\frac{p\mathrm{CO_{2}}^{oc}}{\mu_{\mathrm{CO_{2}}}}\right)},\label{eq:soldim}
\end{aligned}
\end{equation}

where $K_{H}$ is the solubility constant, $p\mathrm{H_{2}O}$ is
the content of water in the ocean, and $\mu$ is the molar mass. The
content of water in the ocean is calculated by assuming the mass of
the Earth's ocean $M_{oc}$.

While sea weathering is a sink for the atmosphere-ocean reservoir,
it behaves like a source for the basaltic crust. Once C is sequestered
into the basalt layer, it starts getting buried by subsequent melt
deposited at the top. The C subsides until it reaches the bottom of
the basalt layer, at which point it is delaminated into the molten
mantle layer. To keep all the quantities in the same metrics, we use
C and CO$_{2}$ interchangeably knowing that C resides in rocks, while
CO$_{2}$ is gaseous form (degassing from mantle).

The rate at which CO$_{2}$ in the basalt is foundered or delaminated
into the mantle is

\begin{equation}
F_{found}=p\mathrm{CO_{2}}^{bas}\frac{v}{d_{bas}} \label{found}.
\end{equation}

Therefore the carbon-dioxide source for the mantle is the carbon dioxide
foundered from the basalt layer, and the sink is the flux being outgassed
through volcanism. Thus, the equations describing this system are

\begin{equation}
\begin{aligned}
\frac{d}{dt}p\mathrm{CO_{2}}^{atm+oc} & =D\left(p\mathrm{CO_{2}}^{man}\right)-W_{sea}\left(p\mathrm{CO_{2}}^{atm},T\right)\\
\frac{d}{dt}p\mathrm{CO_{2}}^{man}      & =F_{found}\left(p\mathrm{CO_{2}}^{bas}\right)-D\left(p\mathrm{CO_{2}}^{man}\right) \\
\frac{d}{dt}p\mathrm{CO_{2}}^{bas}       & =W_{sea}\left(p\mathrm{CO_{2}}^{atm},T\right)-F_{found}\left(p\mathrm{CO_{2}}^{bas}\right) \\
\label{eq:dim}.
\end{aligned}
\end{equation}

Given that the total content of CO$_{2}$ for the planet is fixed
$p\mathrm{CO_{2}}^{atm+oc}+p\mathrm{CO_{2}}^{man}+p\mathrm{CO_{2}}^{bas}=p\mathrm{CO_{2}}^{tot}$,
one of the three equations is redundant. 

The equilibrium carbon dioxide
values for the mantle, atmosphere and ocean are
\begin{align}
p_{E}^{man}=W_{sea}^{E}\frac{V_{man}}{fvA_{p}}  \\
p_{E}^{ao}   =p\mathrm{CO_{2}}^{tot}-W_{sea}^{E}\left(\frac{d_{bas}}{v}+\frac{V_{man}}{fvA_{p}}\right) \\
\end{align}
and $p_{E}^{oc}=p_{E}^{ao}-p_{E}$ in dimensional form.

For typical values refer to Table 1. To be able to solve the system of equations \ref{eq:dim}
or A9, we need to specify the equilibrium conditions
for the atmosphere, namely the equilibrium partial pressure of CO$_{2}$,
$p_{E}$, that sets the equilibrium atmospheric temperature. While
\citet{Menou:2015} took $p_{E}=330$ $\mu$bar , the pre-industrialization
carbon dioxide partial pressure as the equilibrium value for the
Earth, \citet{Haqq-Misra:EBM} argued that a more appropriate value
for an abiotic Earth would be $p_{E}=0.01$ bars, by including all
the carbon dioxide presently stored in the soil. We also take the
equilibrium partial pressure to be $p_{E}=0.01$ bars.

This value for the atmosphere, the solubility constant for the ocean,
and the amount of water in an Earth's ocean, set the equilibrium CO$_{2}$
value for the atmosphere-ocean system via Eq. \ref{eq:soldim} to
be $p_{E}^{ao}=0.038$ bars. 

We also note that in line with astrophysical studies we use the units
of bars instead of moles, and the conversion factor we use is

\begin{equation}
p\mathrm{CO_{2}}[\mathrm{bars}]=1.019710^{5}\times\mu_{\mathrm{CO_{2}}}\frac{g}{4\pi R_{p}^{2}}\times\mathrm{C} \quad [\mathrm{moles}] \label{mol2bar},
\end{equation}

where $g$ is the planet's gravity and $R_{p}$ is the planet's radius.

\subsubsection{Climate Model}

To obtain the atmospheric temperature of a planet provided the carbon
dioxide partial pressure, we use the energy balance model (EBM) by
\citet{Haqq-Misra:EBM}. They provide a fit to the outgoing longwave
radiation ($OLR$), and top of the atmosphere Bond albedo for four
different type of stars including G and M stars. These quantities
depend on the atmospheric carbon dioxide partial pressure, temperature,
and zenith angle $\mu$. We simplify the model to capture global parameters
by fixing the zenith angle to a value $\mu=0.232$ that would provide
a global surface equilibrium temperature of $T_{E}=288$ K for $p_{E}=0.01$
bars for an Earth-like scenario. The zeroth-order energy balance model
we use equates the planet's thermal radiation to the net insolation

\begin{equation}
OLR(p\mathrm{CO_{2}}^{atm},T)=\frac{S_{\star}}{4}\left(1-A\left(p\mathrm{CO_{2}}^{atm},T\right)\right),\label{eq:ebm}
\end{equation}

where $S_{\star}$ is the solar constant at the planet's location,
and $A$ is the top of the atmosphere albedo. This quantity also depends
on the ground albedo which we take from \citet{WilliamKasting:1997}
and modify it for M stars. For the Earth, the ground albedo is taken
to be a discontinuous function of $T$, where for temperatures less
than $273$ K, the ground is considered to be frozen and exhibit a
high albedo. At temperatures below $263$ K the water in the atmosphere
is assumed to condense out entirely owing to the abrupt transition
of the Earth entering a snowball state. However, a recent study by
\citet{Checlair:Mstar} on planets around M stars suggests that unlike on
Earth, ice coverage would proceed gradually, avoiding sudden snowball
transitions, owing to the special spatial insolation pattern they
receive. In our model, we use both the same albedo function for Earth
as well as a modified version that precludes the snowball state following
\citet{Checlair:Mstar}. To achieve this, we allowed for a gradual
ice coverage as temperatures decrease below 273, retaining 30\% of
the land exposure at 150 K, and allowed for cloud albedo at all temperatures. 

The timescale governing Eq. \ref{eq:ebm} is much faster than the
long geophysical timescale involved in the carbon-silicate cycle of
Eqs. \ref{eq:dim} \citep{Menou:2015}. This means that in our model,
surface temperature is calculated instantaneously from the amount
of $p\mathrm{CO_{2}}^{atm}$ at each timestep in the integration of
Eqs. \ref{eq:dim}.

\begin{figure}
\begin{centering}
\includegraphics[width=0.45\textwidth,trim=0 0 0 0, clip]{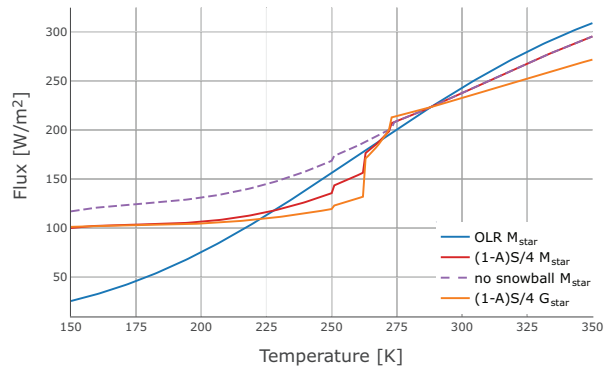}
\par\end{centering}

\caption{Energy Balance Model. Insolation at the location of the habitable
zone (orange) for G stars, (red) M stars and (purple) M star without
snowball states is balanced by the outgoing long-wave radiation OLR
(blue) setting the temperature at the surface for $p\mathrm{CO_{2}}=0.01$
bars. \label{fig:EBM}}

\end{figure}

Figure \ref{fig:EBM} shows the results from the EBM model for the
equilibrium case $p\mathrm{CO_{2}}^{atm}=0.01$ bars. We show the
case of the Earth (orange line) for comparison purposes. Earth's
climate exhibits two stable points, one near 225K where the planet
would be in a snowball state, and the temperate 288K, as well as one
unstable point, near 273K. Furthermore, the snowball state at 225K
is thought to be transient, given that rock weathering is considered
to be suppressed at these temperatures so that outgassing from volcanoes
eventually deglaciates the planet. It is worth mentioning that if
sea weathering can operate at these cold temperatures in a large enough
way as to balance volcanism, this snowball state could be stable on
long term timescales. Sea weathering would have to happen on the ocean
floor below a thick crust of ice, that still allows for carbon dioxide
to diffuse from the atmosphere to the ocean, and volcanism would most
likely have to be sluggish. 

Considering planets around M dwarfs we also modified the EBM to restrict
snowball states. This results in allowing for only one stable temperature
state (purple line).

\subsubsection{Stellar Evolution}

We investigate whether an adequate surface temperature can be kept
during the evolution of an M star for billions of years after the
first 1 active billion. Compared to G type stars like our Sun that
brighten over time after hydrogen ignition takes place, M stars get
dimmer. This means that the habitable zone moves in with time, and hence, 
planets that are found in the habitable zone where once too hot.  This
raises issues as to the likelihood of these planets to be desiccated or
wet after the intense EUV/XUV star fluxes.  While earlier studies \citep{TidalVenuses} estimated planets around M dwarves to be 
completely dry, recent works \citep{Bolmont:AO, Schaefer:AO} 
that use a better prescription for the atmospheric loss \citep{Tian:AO},
suggest that some water is retained, making them prospective planets for habitability.

\begin{figure}
\begin{centering}
\includegraphics[width=0.45\textwidth,trim=0 0 0 0, clip]{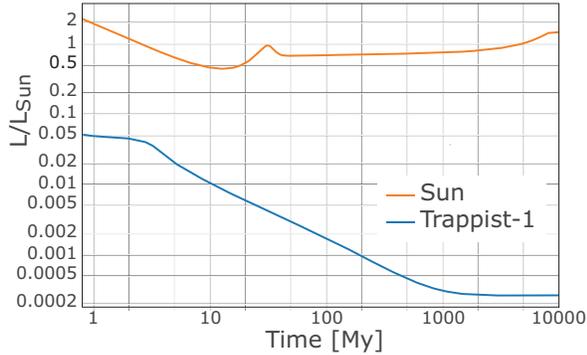}
\par\end{centering}

\caption{Stellar evolution tracks used. The brightness of the M dwarf (blue)
decreases over time, while the Sun (orange) has been brightening since
it reached the main sequence.\label{fig:Stars-evolution}}

\end{figure}

For the evolution of an M dwarf,  we use a spline fit to the model by \citet{Baraffe:2015}
for a star of mass $M_{star}=0.08\,M_{Sun}$ like Trappist-1. Figure
\ref{fig:Stars-evolution} shows the evolution of Trappist-1 compared
to that of our Sun. 

\section{Results}

\subsection{Sea Weathering }

We look first at how an equilibrium may be established and the timescale
associated with it. All the values for the parameters used in the model are
shown in Table 1.

To integrate Eqs. \ref{eq:dim} we can proceed in either of two following ways: 1) determine 
the sea weathering rate needed to ensure the equilibrium of the system is at 
$p_{E}=0.01$ bars and $T_{E}=288$ for the carbon content of the
planet \red{\footnote[2]{ This approach ensures that the
    reference state of the sea weathering in Eq. \ref{eq:Wsea} coincides with the
    equilibrium of the system.}}, and compare this weathering rate to
values obtained for Earth, or 
2) use Earth's estimated weathering rate
at present day conditions of pCO$_2^{atm}=0.01$ and $T= 288$ K \red{to be
the reference state in Eq. \ref{eq:Wsea}}, include the term
for velocity, for other planets, derive the corresponding
equilibrium states, and then
ask if they are suitable for surface liquid water. 

In either case, the functional form of the governing equations remains the same, and from stability analysis (see Appendix)
we conclude that the equilibrium state is in fact stable.

We use both approaches but favour the first one to solve for the equations for two reasons. We use a simple
functional form for sea weathering that could be made more realistic with more understanding (e.g. alkalinity dependence, etc), 
and there is considerable uncertainty behind calculating present day sea weathering rates. Values used previously
range by more than one order of magnitude from 0.225-0.675 Tmol C/year 
\citep{KT_Catling:2017}, 1.75 Tmol C/year \citep{Mills:2014} and 3.4 Tmol C/year \citep{Alt:Wsea, Sleep:2001}. Thus, 
we are not confident enough in our understanding of sea weathering rates on Earth to use it at face value for other planets.

To ensure the equilibrium conditions of $p_{E}=0.01$ bars and $T_{E}=288$
 are at met, the sea weathering at equilibrium must adjust itself to account for the planetary tectonic
conditions and the total carbon dioxide content in the following way

\begin{equation}
W_{sea}^{E}=\left(p\mathrm{CO_{2}}^{tot}-p_{E}^{ao}\right)\frac{fvA_{p}}{\left(V_{man}+fA_{p}d_{bas}\right)}.\label{eq:Wsea_equil}
\end{equation}

\begin{figure}
\begin{centering}
\includegraphics[width=0.45\textwidth,trim=0 0 0 0, clip]{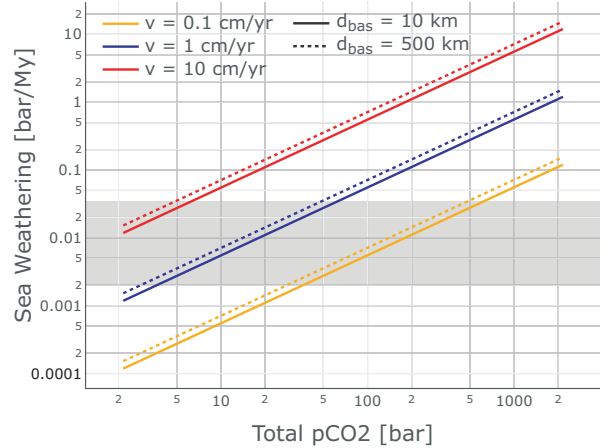}
\par\end{centering}

\caption{Equilibrium Sea Weathering. The rate of uptake of CO$_{2}$ in the
ocean crust needs to adjust to values that depend on the thickness
of the basaltic layer (dash for 10 km and dotted for 500 km) and the
resurfacing velocity (yellow: 0.1 cm/yr, blue: 1 cm/yr, red:10 cm/yr)
for a given total carbon content.
The shaded region shows an estimate on the range of today's present
day values according to \citep{KT_Catling:2017, Mills:2014, Alt:Wsea, Sleep:2001}. 
\label{fig:Equilibrium-Sea-Weathering}}

\end{figure}

If this sea weathering rate can be achieved, and there are no
limiting processes hindering sequestration (i.e. reaction kinetics), then 
equilibrium conditions suitable for water on the planet's surface are 
possible. Figure \ref{fig:Equilibrium-Sea-Weathering}
shows how the sea weathering at equilibrium varies as a function of
total carbon dioxide content for different resurfacing velocities
and basaltic crust thicknesses. 


We consider the total planetary carbon content to be the same as for Earth estimated
at $2.5\times10^{22}$ moles C/yr\citep{Sleep:2001} or $215.51$
bars. Io's estimated resurfacing rate is 0.55-1.06 cm/yr \citep{Io:Book}
to a few cm/yr \citep{Io:resurfacing} and crustal thickness is $14$
km$-30$ km at a minimum \citep{Jaeger:IoPlate,Carr:Io}. We take
as our fiducial case a resurfacing velocity of $v=1$ cm/yr and basaltic
crust layer thickness $d_{bas}=10$ km, in alignment with Io, which yield an equilibrium
sea weathering rate of $120$ bars/Gy. This is above the range 
estimated for present day Earth ($1.9 - 29 $ bar/Gyr). For comparison, a less active
planet, with a resurfacing rate of $v=0.1$ cm/y, would require equilibrium
sea weathering rates of $12$ bar/Gy, well within present day Earth's estimates.

It is clear that the most important
factor is the carbon content of the planet and the resurfacing velocity.
The shaded region shows present day estimates for sea weathering rate
on Earth.

If there is a limit to how much sea weathering can take place, 
for a given total carbon content then lower resurfacing velocities are needed 
in order to have an equilibrium state, while the thickness of the basaltic layer is less important
(see Fig. \ref{fig:Equilibrium-Sea-Weathering}). 
In turn, resurfacing velocities depend on the tidal heat flux, or tidal forcing. Thus,
if there is a limit to sea weathering, there will be a limit
to tidal heating above which the planet will not exhibit an equilibrium
atmospheric state over long timescales that favours liquid water. 
It is beyond the scope of this paper to calculate the exact limits, as this
would require building a thermal evolution model for the planet.

Likewise, for a given resurfacing rate set by tidal heating, planets
with less amount of C are favoured in maintaining surface habitable conditions 
via tidally-induced tectonism.  

If instead, we take Earth sea weathering value
$W_{sea}^{\oplus}$ \red{at face value and valid only in the reference
  state},
we can calculate the equilibrium conditions for atmospheric
temperature and CO$_2$ pressure \red{for other planets}. \red{ For this case we make explicit the
dependence on the velocity by setting $\omega=
\frac{v}{v_{\oplus}}$ and take $v_{\oplus}= 3$ cm/y. 
We obtain the equilibrium conditions for a given set of resurfacing velocities, basaltic layer thicknesses and total carbon content
by solving for the value of $p_E$ that satisfies the equation }

\begin{equation}
p_\mathrm{CO_{2}}^{tot} =  p_{man}^{E} \left( 1 + \frac{f d_{bas} A_{p}} {V_{man}}  \right) + p_{E} + p_\mathrm{CO_{2}}^{oc} (p_{E})  
\end{equation}

given that,

\begin{equation}
p_{man}^{E} = \frac{V_{man}}{fvA_{p}} W_{sea}^{\oplus} \frac{v}{v_{\oplus}} \left(p_E/p_\oplus \right)^{\alpha} 
        \exp\left(\frac{E}{R}\left(\frac{1}{T_{\oplus}}-\frac{1}{T_E}\right)\right)    
\end{equation}
        
\begin{figure}
\begin{centering}
\includegraphics[width=0.6\textwidth,trim=0 0 0 0, clip]{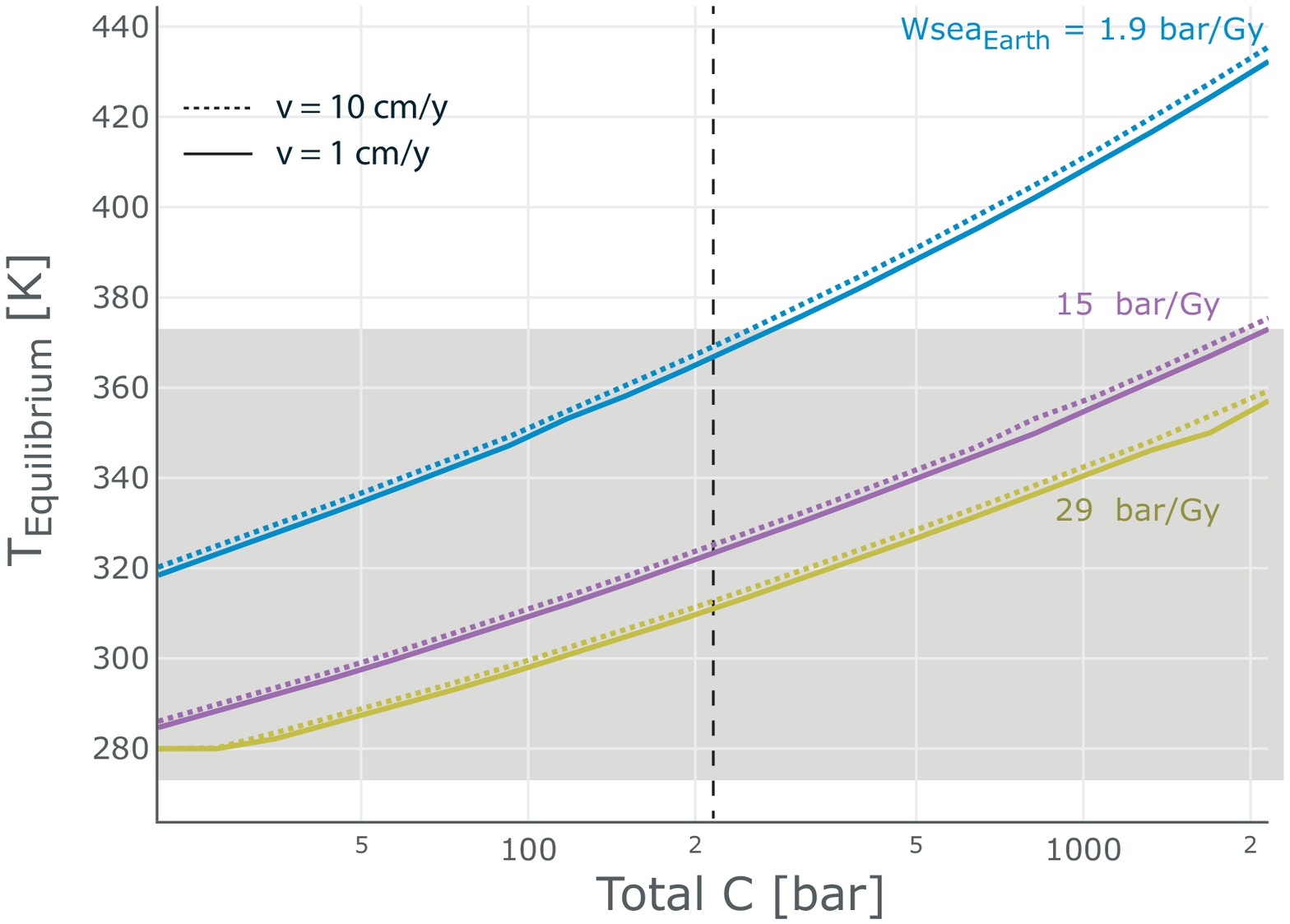}
\par\end{centering}

\caption{Equilibrium surface temperature for planets with different C contents
provided three sea weathering rates estimates for the Earth, blue: 1.9  \citep{KT_Catling:2017} , 
purple: 15 \citep{Mills:2014}, and green: 29 bar/Gy \citep{Sleep:2001}, 
and spreading rate of $v_\oplus = 3$ cm/y. Dashed line corresponds to the estimated C content for the Earth according to \citet{Sleep:2001}.
The shaded region shows conditions for liquid water
provided a 1 bar atmosphere. 
\label{fig:Tnew}}

\end{figure}

Figure \ref{fig:Tnew} shows the range of equilibrium temperature values as a function of different planetary carbon contents, for an Earth's average spreading
rate of $v_\oplus = 3$ cm/y and three different estimates of Earth's sea weathering rate: 
$W_{sea} ^ {\oplus} = 0.55$  Tmol/y = 1.9 bar/Gy \citep{KT_Catling:2017} ,
$W_{sea} ^ {\oplus} = 1.75$  Tmol/y = 15 bar/Gy \citep{Mills:2014} , and
$W_{sea} ^ {\oplus} = 3.4$  Tmol/y = 29 bar/Gy \citep{Sleep:2001}. 
It can be seen that planets with modest amounts of total carbon can achieve
habitable conditions easier than those that have more carbon content. For example, planets with carbon contents a 
few times that of the Earth can have liquid water (shaded region) when assuming a 1 bar atmosphere, albeit at hotter
conditions.

Either way of calculating sea weathering, it is clear that planets with low carbon contents, and/or sluggish resurfacing
velocities are favoured for exhibiting a carbon-silicate cycle that keeps the atmospheric temperature at habitable conditions.

\subsection{Equilibrium Timescale }

Having established the equilibrium conditions, we proceed to solving the governing equations. 
Our approach is to set the sea weathering rate to a value that allows
for an equilibrium similar to the Earth with $p_E = 0.01 $ bar and
$T_E = 288$K. 
However, the behaviour is expected to be qualitatively similar had we
used fixed the sea weathering rate to the Earth's value given the same functional form of the Equations.

We first start with a fixed present day Sun to find out how long it
takes to reach the equilibrium state and how it changes with different
parameters. As initial conditions we tested different ones (similar
to those used by \citet{Foley:Area}) and show only the most extreme 
case of disequilibrium where there is no carbon in the mantle to begin
with. While this case is not really realistic it stand to show that the system
can recover equilibrium even for these extreme beginnings.
We consider the cases where a) these planets reach snowball
states and b) where they do not. We vary the initial atmospheric CO$_{2}$
values to allow for cold or hot beginnings. 

\begin{figure}
\begin{centering}
\includegraphics[width=0.45\textwidth,trim=0 0 0 0, clip]{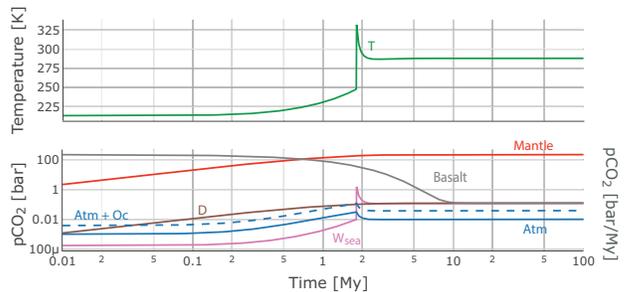}
\par\end{centering}

\caption{CO$_{2}$ evolution towards equilibrium for tidally heated planets
in the habitable zone. Temperature (top, green) changes as the partial
pressure of atmospheric CO$_{2}$ (blue) evolves. The CO$_{2}$ content
of the atmosphere+ocean reservoir (dashed blue) evolves commensurate
with the atmospheric CO$_{2}$ partial pressure given an assumed solubility
independent of temperature. The C content in the basalt (grey) and
mantle (red) reach the equilibrium state at a few tens of My once
the sea weathering rate (pink, in bar/My) and volcanic degassing
(brown, in bar/My) reach
parity. \label{fig:Sfix_p0.001} }
\end{figure}

Figure \ref{fig:Sfix_p0.001} shows how the system reaches equilibrium
starting from a cold state. With no C in the mantle, the system initially
starts with no outgassing into the atmosphere. With cold temperatures
($T=215.5$ K for $p_{0}^{atm}=0.001$), the weathering rate is very
small and draws down little amount of CO$_{2}$ from atmosphere (and
ocean) into the basaltic layer that starts as a rich C reservoir.
Foundering from this basaltic crust slowly starts building the mantle's
C reservoir, as outgassing is outpaced. This little outgassing slowly
builds up more CO$_{2}$ in the atmosphere, so that it heats up and
starts melting the ice on the surface. At some point there is enough
CO$_{2}$ in the atmosphere that the planet deglaciates completely,
this changes the albedo suddenly to much lower values and the planet
transitions into a hot state (the only permanent stable point in the EBM plus evolution 
equations is at high temperatures). The system overshoots from the equilibrium
point because too much CO$_{2}$ had built in the deglaciation phase. This overshoot is  
controlled by the rate of outgassing and the time it has taken the planet to deglaciate.
At this very hot state, the sea weathering rate increases by an order
of magnitude and quickly draws back down the excess CO$_{2}$ in the
atmosphere, all the while decreasing C in the basaltic crust and increasing
it in the mantle, relaxing into the equilibrium state.  Analogous behaviour
has been discussed in the context of Earth and deglaciating from previous
snowball states \citep{Hoffman:snowballEarth}.

\begin{figure}
\begin{centering}
\includegraphics[width=0.45\textwidth,trim=0 0 0 0, clip]{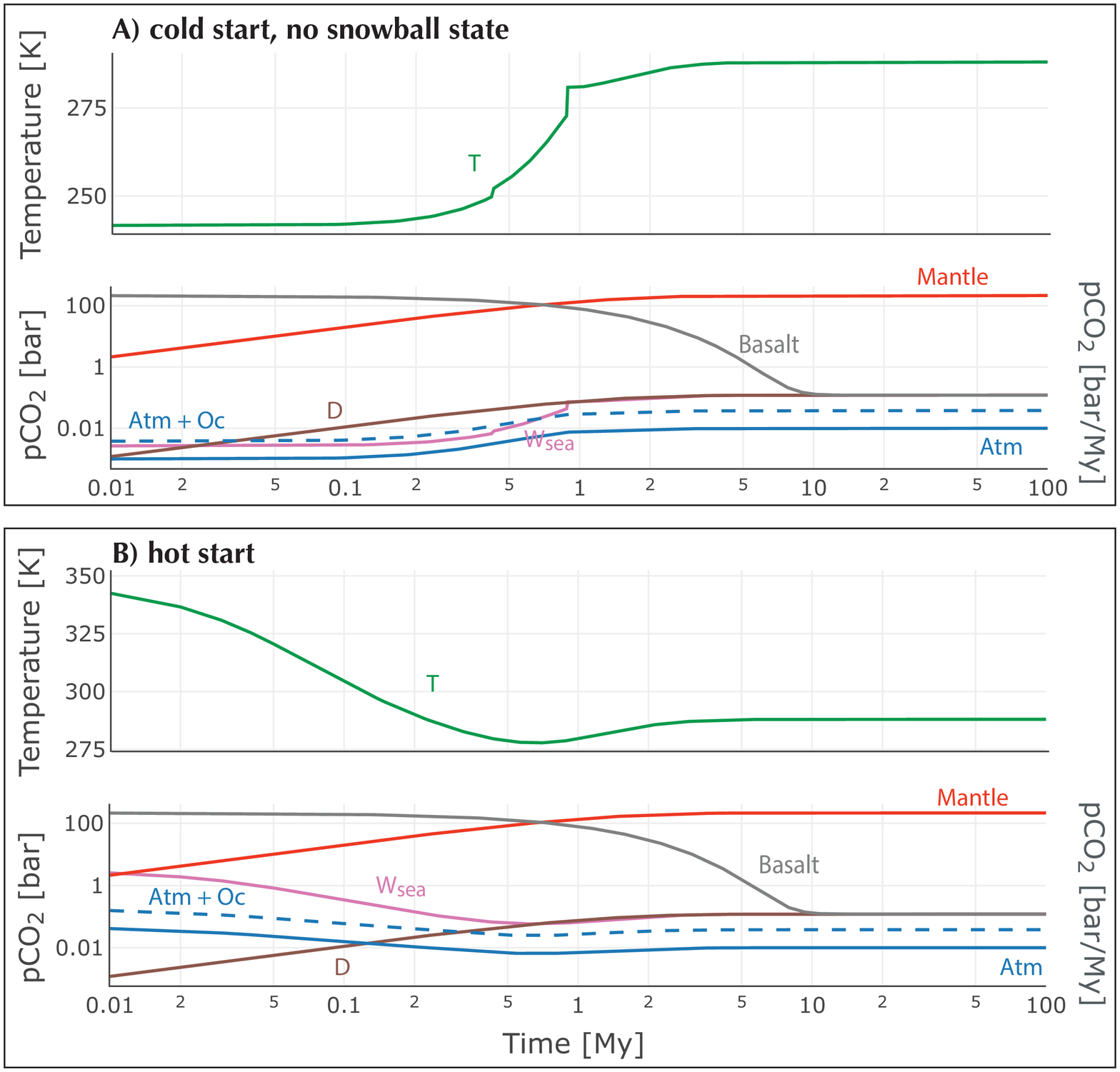}
\par\end{centering}

\caption{CO$_{2}$ evolution towards equilibrium for tidally heated planets
in the habitable zone. Top) Cold initial state for planets without
snowball states. Bottom) Hot initial state.\label{fig:Sfix_Two}}

\end{figure}

If we restrict the ice coverage of the planet to eliminate snowball
transitions, following \citet{Checlair:Mstar}, then the inital state
is less cold ($T=243$ K) and no overshooting happens. See Fig. \ref{fig:Sfix_Two} top panel.
The planet just heats up until it reaches the equilibrium state. Discontinuities
in the temperature come from discontinuities in the ground albedo
as a function of surface temperature via the amount of land, snow
and ice coverage that have been modelled in a simple fashion. 

The last case we considered was a hot beginning with $p_{0}^{atm}=0.05$
bars (the hottest point allowed in the EBM model for M stars). The
evolution from a snowball or no-snowball planet is the same. For this
hot beginning, sea weathering starts at a high rate, drawing down
CO$_{2}$ from the atmosphere, bringing down the surface temperature.
In the meantime, the basaltic layer is foundering more C to the mantle
than the mantle degasses, so that the mantle's reservoir builds up.
The system also slightly overshoots, but not to a point of glaciation,
and then relaxes into the equilibrium state. 

We find that the timescale to reach equilibrium is about 10 -100 My and
is independent of total CO$_{2}$ content, activation energy $E_{sea}$ values, 
or dependency of the sea weathering rate on the amount of atmospheric CO$_{2}$, $\alpha$. 
The factors that change this timescale are the crustal thickness $d_{bas}$, resurfacing velocity $v$,  and outgassing
rate. An order of magnitude increase in crustal thickness from 10
to 100 km increases the timescale by about one order of magnitude
from $\sim 10$ to $\sim100$ My. An increase in resurfacing velocity from 1 to 10 cm/yr
decreases the timescale by about one order of magnitude. A decrease in outgassing fraction by one order of magnitude
from $f=0.1$ to $0.01$, increases the timescale by a factor of a few.

While these relationships were obtained through parameter exploration, we also
obtained an expression for the equilibrium timescale by
non-dimensionalizing \ref{eq:dim} (see Appendix). The expression is
\begin{equation}
\tau \simeq \frac{d_{bas}}{v \sqrt{f}}
\end{equation}

On the other hand, larger values of total carbon inventories while not
changing the timescale to reach equilibrium, may affect the timing
of complete deglaciation (by a few My) while preserving the amount
of overshoot in atmospheric CO$_{2}$ , and thus the atmospheric temperature
(to about $330$ K). Different values of $v$ have the same effect
on changing the timing of the overshoot, but not the peaks and troughs
in temperature. Thus, for reasonable values for the parameters in
the model, the timescale to reach equilibrium is 10-100 My. Therefore, shorter
or longer time scale processes are not expected to affect the planet's
ability to sustaining equilibrium.

\subsection{Long-Term Evolution }

For completeness, along the same lines, we also looked at how a tidally-heated
planet may evolve as the M star dims over time. Because the evolution
of stars changes in a billion year timescale, we find the planets
reaching equilibrium at 10-100 My, as expected. Thus, we only show
the evolution case of a hot start as it is likely that planets that
end up in the habitable zone around a M stars, started hot (unless
there is some mechanism for migration that might have brought them
from further out,  a la \citet{Hansen:2012} ). To stay within the bounds of the EBM model by \citet{Haqq-Misra:EBM}
our initial start is at $p_{0}^{atm}=0.033$ bars. After 10-100 My
the CO$_{2}$ contents of the reservoirs reach the equilibrium state
for the solar luminosity at the time. Given that the star is brighter
in the past, the CO$_{2}$ content is lower ($p^{atm}=0.007$ at 2.9
By ago), while the atmospheric temperature ($T=289$ K) is slightly
larger than present day. Our choice for starting the evolution 3 By
ago is tied to the range in which the EBM model is valid. However,
there is no reason to believe that this mechanism would not operate
further into the past at larger insolation values. The limit would
be set by reaching the runaway greenhouse. 
In other words, planets in the habitable zone of M dwarfs may start hot, with a steam atmosphere, that partially evaporates, while
the star is active. As luminosity decreases, the remaining water in the atmosphere will condense out \citep{Bolmont:AO, Schaefer:AO}, or perhaps
any water stored in the mantle can outgas to the surface. For planets located in near-resonant chains, which can keep eccentricities from decaying to zero, tidally driven tectonics may start taking place.

Our simple model shows that tidally-locked planets may have a built-in
mechanism to regulate the amount of CO$_{2}$ in the atmosphere to
allow for long term stability of surface liquid water, similar to
Earth. However, unlike Earth's carbon-silicate cycle that relies on
plate tectonics, these tidally heated planets recycle material vertically
via continuous volcanism and foundering. 

\begin{figure}
\begin{centering}
\includegraphics[width=0.45\textwidth,trim=0 0 0 0, clip]{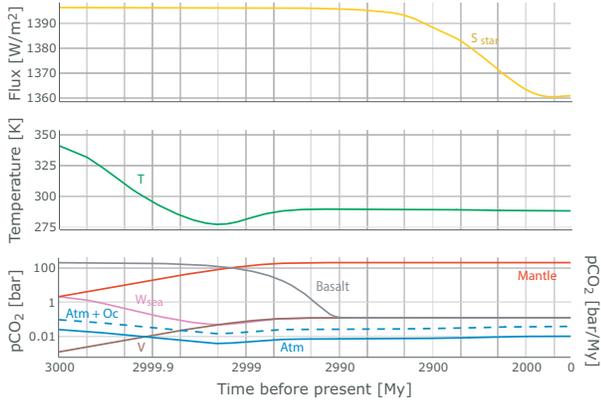}
\par\end{centering}

\caption{CO$_{2}$ long-term evolution as the host M star evolves form an initial
hot state}

\end{figure}

\subsection{Sub-Earths and Super-Earths}

A simple extension of this work is to consider planets that are less or more
massive than Earth, while still being rocky. Low-mass exoplanets,
including super-Earths are now known to be common in our galaxy \citep{Howard:2010}. 
We consider planets that have the same major element composition as Earth, including the
same core mass fraction and assume that these planets experience constant volcanism.
We use parameters for planetary radius, gravity and core radius to account
for the different mass according to the internal structure model by \citet{Valencia:2006} 

In terms of the planetary carbon inventory we consider two scenarios: 1) planets with same total C content as  
Earth , and 2) planets with the same C content per unit mass. In reality, due to the volatile character of carbon, we do not fully understand how 
Earth acquired the amount it has, and how to extrapolate this accurately to other terrestrial planets. Thus our two scenarios may be thought of
as possibilities, from which we can draw \red{a few} conclusions. 

We use the same resurfacing velocity and crustal thickness for all
planets, independent of their mass \red{as a first order approach.} 
The timescale to reach equilibrium remains unchanged, although in the case
of cold initial states with snowball transitions, the overshoot is
delayed for larger planets. Thus, \red{when considering only the effects of
geometry and gravity}, our proposed mechanism {is robust}.

\begin{figure}
\begin{centering}
\includegraphics[width=0.45\textwidth,trim=0 0 0 0, clip]{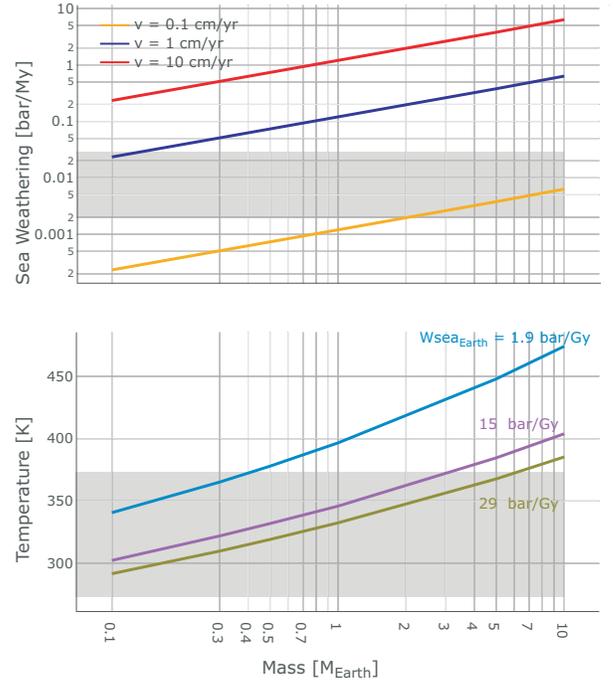}
\par\end{centering}
\caption{Tidally driven tectonics in sub and super-Earths with scaled carbon content. Top: Sea weathering rates needed to 
ensure equilibrium of $T=288 $K and pCO$_2=0.01$ bars for planets of different masses and corresponding carbon contents. 
Bottom:  Equilibrium surface temperatures if sea weathering rates are the same as the Earth given three different estimates ( same 
as in Fig \ref{fig:Tnew}) \label{fig:SE} }
\end{figure}

Not surprisingly, when we allow planets' carbon inventory to scale with mass, it affects the system's requirements. If we require an equilibrium around T=288K, planets with lower
carbon contents require more reasonable sea weathering rates (Fig. \ref{fig:SE} top). If instead, we take Earth's sea weathering at face value, smaller planets
with lower carbon contents experience equilibrium temperatures that are temperate whereas large planets would be too hot (Fig. \ref{fig:SE} top).  

In general, in terms of a negative feedback mechanism that regulates surface temperature via CO$_2 $ sequestration, 
there needs to be a balance between outgassing that depends on planetary carbon content and weathering. Thus, lower planetary carbon contents are favoured.

Another aspect of planetary mass is how it affects the ability of the planet to advect melt to the surface given the fact that there is a crossover density at some pressure below which any melt produced is negatively buoyant and stays within the mantle \citep{Ohtani:melt}. The depth at which this occurs depends on the gravity, and thus on planet mass. A simple gravity scaling based on \citet{Valencia:2006}, yields $g \simeq g_{E} \left( M/M_{E}\right)^{1/2}$, which translates to a depth scaling of $h \simeq h_{E}\left( M/M_{E}\right)^{-1/2}$. Thus, smaller planets are more likely to produce melt above the depth at which the crossover density happens.  Following \citet{Sakamaki:melt} and this simple scaling (which ignores compression effects) we obtain for a $10\,M_E$ planet (with the same core-mass fraction as Earth) depths for the crossover density of 110 km, 130 km, and 450 km for anhydrous magmas, hydrated magmas at the 2\% and 8\% level, respectively. For a $0.1\,M_E$ planet the corresponding depths are 1100 km, 1300 km, and beyond the core-mantle boundary of the planet.  Thus, details about the composition of the mantle and melt as well as where tidal dissipation takes place are needed to establish exactly which planets will produce melt that can be advected to the surface, and sustain the mechanism we propose. These effects are beyond the scope of this paper, and left for future work. However, qualitatively, melt is more easily advected in smaller planets, with hydrated magmas, and thus making small wet planets more suitable to exhibit habitability from tidally heated tectonism.

\begin{table*}
\caption{Model Parameters}

\begin{center} 
\begin{tabular}{ c|c|c|c  }  
Parameter & symbol & value & Reference \\ 
\hline
Sea weathering dependence on CO$_2$ & $\alpha$ & $0.23$ & (1) \\ 
Sea weathering at equilibrium & $W_{sea}^E$ & $0.12$ bar/My & calculated \\   
Equilibrium atmospheric temperature & $T_E$ & $288$ K & assumed \\ 
Equilibrium atmospheric CO$_2$ partial pressure & $p_E$ & 0.01 bars & (2) \\
Activation Energy & $E_{sea}$ & 41 kJ & nominal value, (1) \\
Outgassing CO$_2$ fraction & $f$ & 0.01 & nominal value, assumed \\
CO$_2$ molecular weight & $\mu_{\mathrm{CO}_2}$ & $44.1$ mol/g & calculated \\
H$_2$O molecular weight & $\mu_{\mathrm{H_2O}}$ & $18.015$ mol/g & calculated \\
Mass Ocean & $M_{oc}$ & $1.4 \times 10^{24}$ kg &  \\ 
Solubility constant & $K_h$ & $235.48$ bar & calculated \\
Planet Radius & $R_p$ & $6371$ km & 3 \\
Core's Radius & $R_c$ & $3400$ km & 3 \\
Planet Mass & $M_p$ & $5.972 \times 10^{24}$ kg & 3 \\ 
Zenith angle & $\mu$ & 0.232 & calculated \\
Sun's luminosity today  & & 1361 W/m$_2$ &  \\ 
Basalt crust thickness & $d_{bas}$ & $10$ km & nominal value, assumed \\
Resurfacing velocity & $v$ & $1$ cm/y & nominal value, assumed \\
Total carbon content & $p\mathrm{CO_2}^{tot}$ & $215.51$ bars & 4 \\
Characteristic  timescale &  $\tau$ & $3$ My & calculated \\
\hline 
\end{tabular} 
\end{center}

\tiny

Refs: (1) \citet{BradyGislason:1997}, (2) \citet{Haqq-Misra:EBM},
(3) \citet{PhysicsEarth}, (4) \citet{Sleep:2001}

\end{table*}
\normalsize

\section{Discussion}

The simple model we propose has three elements at its core that make
for the built-in thermostat: 1) the sea weathering rate depends sensitively
on the atmospheric temperature (via controlling the temperature of
the deep ocean), 2) the atmospheric CO$_{2}$ can be drawn out and
sequestered into a reservoir when needed, and 3)
this reservoir is connected back into releasing CO$_{2}$ into the
atmosphere via outgassing, which in our proposed model happens via
foundering of the basaltic layer into the mantle that, in turn, continuously
outgases into the atmosphere by advecting melt onto the surface. 

This model does not consider possible limits to the weathering rate.
By construction we have invoked a mechanism that continuously exposes
fresh basaltic crust available for weathering, in analogy to Earth. 
However, if volcanism is too infrequent on timescales that are longer
than $\sim$10-100 My, it could be a problem for the system to maintain
equilibrium. \citet{BradyGislason:1997} noted that seafloor weathering
on Earth seems to occur mostly within the first 3 My and stops after
10 My after crust production. Thus, for continuous sea floor weathering,
enough volcanism should happen at most every $\sim$million year. 

Also, incipient volcanism may limit the amount of weatherable material
in a similar way to the transport-limited scenario explored by \citet{West:2005,West:2012}
for Earth. In the transport-limited regime, the replenishment of fresh
Ca and Mg ions is the bottle neck to weathering. In the case of the
Earth it can be due to low erosion rates \citep{West:2005,West:2012}.
In rocky exoplanets it can be due to limited land exposures \citet{Foley:Area}.
In our case it can be due to limited volcanism.  Including a transport-limited
sea weathering rate, would impose a limit to the amount of total planetary carbon, and/or
a resurfacing rates (which are dependent on tidal heating values) 
below which the planet has the ability to sequester C at a high enough rate 
to keep CO$_2$ atmospheric values below a greenhouse atmosphere.  
Thus, understanding better what parameters control sea weathering
would help us determine how ubiquitous this kind of C-Si cycle can
be in exo-Earths. 

In addition, we treated seafloor weathering in the simplest way possible.
For example, our equation for seafloor weathering lumps alkalinity
into the surface temperature dependence, and thus
we omit a treatment for ocean chemistry \citep{KT_Catling:2017}.
These improvements are left for future work to bring focus to the
main ingredients laid out in this study.

Our proposed scenario excluded the existence of rock weathering on
continents. However, it may be that orogeny can happen as a secondary process
to tidally induced volcanism as suggested on Io (\citet{Io:BookChapter}
and references therein). If so, weathering may proceed in both the 
continents and on the seafloor. 

We note that in our modelling we have taken the parameters for the thickness 
of the basaltic layer and the resurfacing velocity as independent quantities, while in reality 
they are connected via the contribution of tidal heating to the total heat flux of the planet
(factor $a$ in Eq. \ref{d_bas} ).  We have taken this route because to properly assess this contribution
one would have to model the thermal history of the planet including 
an accurate description of the tidal dissipation parameter $Q$, and the effect of melt on it, which is
beyond the scope of this study. 

Another refinement to our simple model may be to include the effects of phase transitions happening
within the basaltic layer.  On Earth, the basalt to eclogite transition may cause delamination within a thick crust
and cool the surrounding mantle faster than otherwise, and limit the size of the basaltic crust.
Including this effect on our model would require adding another layer between the basaltic crust and the mantle 
from which outgassing proceeds.  Because of mass balance, adding another layer would not change the 
character of the equations, and thus the results presented. 

Future work may be extended to include thermal history calculations, sweeping of parameter 
space for planets at different orbital configurations,  different
carbon contents, and include the effects of transport-limited sea weathering via limited volcanism. 

If our proposed mechanism to regulate atmospheric CO2 takes place in planets like Trappist-1$d$
the CO$_{2}$ content in the atmosphere would be around values that
would yield liquid water on its surface, namely $\sim0.5$ bars, assuming
the planet is abiotic. Indeed, if there is a negative feedback mechanism 
that enables a thermostat taking place in other
planets, enabled by either the mechanism proposed here or by plate
tectonics \red{, or even perhaps in stagnant lid \citep{Foley:2017}}
the CO$_{2}$ content of the atmosphere has to be commensurate
with insolation values, something we can test for with enough atmospheric
data.  However, because planets \red{ that are substantially tidally
  heated most likely are getting rid of heat via heat piping, 
we can be guided by estimates of the tidal dissipation from orbital dynamics to pinpoint whether or not
we expect heat piping to occur instead of plate tectonics or stagnant lid.}

\section{Summary}

In summary, we propose a new mechanism for rocky planets around M
dwarfs to have a climate-controlling feedback mechanism that can keep
liquid water stable for billions of years. An analogous carbon-silicate
cycle can operate on planets by recycling carbon between the atmosphere,
a basaltic crust and the mantle via tidally-induced volcanism, basaltic
formation and sea weathering, plus foundering. In contrast to plate
tectonics that enables the carbon-silicate cycle on Earth, these planets
would achieve the same principle by recycling material vertically. 

Continuous exposure of basaltic crust through volcanism can be weathered
to sequester C from the ocean and atmosphere, to draw down CO$_{2}$
atmospheric levels when values are too high, or outgases CO$_{2}$
from the mantle to replenish atmospheric values when they are too
low, as long as sea weathering depends on atmospheric temperature. 

Therefore, the tidal properties of planets around M dwarfs, in the
absence of plate tectonics, may enable stable climates suitable for
habitability, making the search for these planets more attractive
than it already is. 

\red{The equilibrium timescale of $\sim$ 10-100 Myr underlying this mechanism ($\tau
\simeq d_{bas}/v\sqrt{f}$) is very different from the timescales for
stellar brightness evolution or flares, thus remaining impervious to
these changes}.

This mechanism may be tested by retrieving atmospheric CO$_{2}$ values
from nonzero eccentric planets in the habitable zone that are dissipating to much heat via
pipe heating to exhibit plate tectonics. If this type
of tectonism is happening and controlling climate, the values for CO$_{2}$
should be consistent with insolation values. Trappist-1 planets in
the habitable zone may be an example.

\acknowledgements{}
The authors thank Yanqin Wu and Dan Tamayo for useful discussions. We
thank Brad Foley for his thoughtful review that help improve this manuscript. VT and ZZ would like to thank the Centre for Planetary Sciences for
the undergraduate research fellowships. This work was in part funded
by NSERC grant RBPIN-2014-06567.

\appendix
\section{Linear Stability and Equilibrium Timescale}

This system of ordinary differential equations that govern this system (Eq. \ref{eq:dim}) can be non-dimensionalized
to yield

\begin{equation}
\begin{aligned}
\frac{d}{dt^{'}}p\mathrm{CO_{2}^{'}}^{atm+oc} & = \frac{W_{sea}}{p_{E}^{ao}} \tau \left( p\mathrm{CO_{2}^{'}}^{man}-\left(p\mathrm{CO_{2}^{'}}^{atm}\right)^{\alpha}\exp\left(\frac{E}{R} \left(\frac{1}{T_{E}}-\frac{1}{T}\right)\right) \right)  \\
\frac{d}{dt^{'}}p\mathrm{CO_{2}}^{man} & = \tau \frac{fA_{p}v^{2}}{V_{man}d_{bas}}\left(\frac{p_{E}^{ao}}{W_{sea}^{E}}\right)  \left(1-p\mathrm{CO_{2}^{'}}^{atm+oc}\right)+
\tau \left(\frac{fvA_{p}}{V_{man}}+\frac{v}{d_{bas}}\right) \left(1-p\mathrm{CO_{2}^{'}}^{man}\right) \\
\label{eq:nondim1}
\end{aligned}
\end{equation}

with nondimensional variables 
\begin{align}
p\mathrm{CO_{2}^{'}}^{atm+oc}=p\mathrm{CO_{2}}^{atm+oc}/p_{E}^{ao} \\
p\mathrm{CO_{2}^{'}}^{atm}=p\mathrm{CO_{2}}^{atm}/p_{E} \\
p\mathrm{CO_{2}^{'}}^{man}=p\mathrm{CO_{2}}^{man}/p_{E}^{man} \\
p\mathrm{CO_{2}^{'}}^{oc}=p\mathrm{CO_{2}}^{oc}/p_{E}^{ao}  \\
t^{'}=t/\tau 
\end{align}

After inspection of the numerical results solving the dimensional
equations, as well as combining both equations in \ref{eq:nondim1}
into a second order differential equation for
$p\mathrm{CO_{2}^{'}}^{man}$ and retaining the non derivative terms,
we conclude that the most appropriate timescale is 
\red{
\begin{equation}
\tau = \frac{1}{v}\sqrt {\frac{d_{bas}V_{man}} {fA_{p}} } ,
\end{equation}
}

which in the limit of a thin basaltic crust layer $d_{bas}/R_p \ll 1$ becomes

\begin{equation}
\tau = \frac{d_{bas}} {v} \frac{1}{\sqrt{f}}. 
\label{eq:tau}
\end{equation}

With this choice for non-dimensionalisation the equations become: 

\begin{align}
\frac{d}{dt^{'}}p\mathrm{CO_{2}^{'}}^{atm+oc} & = R_{1} p\mathrm{CO_{2}^{'}}^{man}-\left(p\mathrm{CO_{2}^{'}}^{atm}\right)^{\alpha}\exp\left(\frac{E}{R}\left(\frac{1}{T_{E}}-\frac{1}{T}\right)\right)\nonumber \\
\frac{d}{dt^{'}}p\mathrm{CO_{2}}^{man} & =\frac {1}{R_1} \left(1-p\mathrm{CO_{2}^{'}}^{atm+oc}\right)+R_{2}\left(1-p\mathrm{CO_{2}^{'}}^{man}\right) \\
\label{eq:nondim}
\nonumber 
\end{align}
and

\begin{equation}
p\mathrm{CO_{2}^{'}}^{atm+oc}=p\mathrm{CO_{2}^{'}}^{oc}\left(1+\frac{\frac{K_{H}}{\mu_{\mathrm{CO_{2}}}}}{\frac{p\mathrm{H_{2}O}}{\mu_{\mathrm{H_{2}O}}}+\frac{p\mathrm{CO_{2}}_{E}^{atm+oc}}{\mu_{\mathrm{CO_{2}}}}p\mathrm{CO_{2}^{'}}^{oc}}\right),\label{eq:solnondim}
\end{equation}

where the dimensionless groups related to this problem are

\begin{align}
R_{1} & =  \frac{W_{sea}^{E}}{p_{E}^{ao}} \frac{1}{v} \sqrt{ \frac{d_{bas} V_{man}}{f A_p}}  \\
R_{2} & = \left(\frac{fvA_{p}}{V_{man}}+\frac{v}{d_{bas}}\right) \tau,
\end{align}

which in the limit of thin basaltic crust layer become 
\begin{align}
R_{1} & =   \frac{W_{sea}^{E}}{p_{E}^{ao}} \frac{d_{bas}}{v \sqrt{f}} \\
R_{2} & = \left(\frac{v}{d_{bas} }\right)^2 \frac{1+f}{f}.
\end{align}

In non-dimensional form it is easy to see the combination of 
parameters that govern the equation. For example,  $\frac{p_{E}^{ao}}{W_{sea}^{E}} $ come as a block, as well as $\frac{d_{bas}}{v}$ in the thin 
basaltic shell limit.

We are interested in determining if the equilibrium point 
$p\mathrm{CO_{2}^{'}}^{atm} = p\mathrm{CO_{2}^{'}}^{atm+oc} = p\mathrm{CO_{2}^{'}}^{man} = 1$ and
$p\mathrm{CO_{2}^{'}}^{oc} = 1- \frac{p_{E}}{p_{E}^{ao}}$ is stable.

Evaluating the jacobian at this fixed point we obtain

\[ J = \left[ \begin{array}{cc}
- A & R_{1}  \\
- 1/R_{1} & -R_{2}  \end{array} \right] 
 \]

where 
\begin{equation}
A = R_1 \frac{p_{E}^{ao} }{p_{E}} \left( \alpha + \frac{E}{RT_{E}^2}  \frac{d T}{p\mathrm{CO_{2}^{'}}^{atm}} \right) \left(1 - \frac{d p\mathrm{CO_{2}^{'}}^{oc} }{d p\mathrm{CO_{2}^{'}}^{atm+oc} } \right),
\end{equation} 

and 
\begin{equation}
1 - \frac{d p\mathrm{CO_{2}^{'}}^{oc} }{d p\mathrm{CO_{2}^{'}}^{atm+oc} } = \frac{K_H \mu_{CO_2} \mu_{H_2O} p_{H_2O}}{K_H \mu_{CO_2} \mu_{H_2O} p_{H_2O} +
\mu_{CO_2}^2 p_{H_2O}^2 + 2 \mu_{CO_2} \mu_{H_2O} p_{H_2O} p_{E}^{ao} p\mathrm{CO_{2}^{'}}^{oc} + \mu_{H_2O}^2 {p_{E}^{ao}}^2 p\mathrm{{CO_{2}^{'}}^{oc}}^2}, 
\end{equation}, 

a quantity that is always positive.

Because the derivative of temperature with respect to atmospheric carbon dioxide is always positive, then $A$ is always a positive quantity.

Obtaining the eigenvalues of the jacobian matrix we find
\begin{equation}
\lambda_{+,-} = - \frac{1}{2} \left( A + R_{2}\right) \pm \sqrt{ \left( A - R_{2}\right)^2 - 4}.
\end{equation}

There are two possibilities, either the term in the square root is negative, or positive.  If it is negative, given that $A$ and $R_{2}$ $>0$, the real part of the eigenvalues is negative and the equilibrium point is a stable solution even if there is decaying oscillatory behaviour around it.  If the term in the square root is positive, then we have to determine in which cases the eigenvalues are negative. Trivially, $\lambda_{-}$ is always negative.  For $\lambda_{+}$, the condition is that $A+R_2  > \sqrt{\left( R_2 - A \right)^2 - 4}$ yield negative eigenvalues.  As both quantities are non-negative, the condition can be reduced to $A R_2 > -1$ , which is always satisfied.  

Therefore, we conclude that the equilibrium point is stable regardless of what assumptions are made about the amount of tidal heating reflected on values for $v$ and $d_{bas}$.

\bibliographystyle{apj}
\addcontentsline{toc}{section}{\refname}\bibliography{Notes}

\begin{thebibliography}{66}
\expandafter\ifx\csname natexlab\endcsname\relax\def\natexlab#1{#1}\fi

\bibitem[{{Alt} \& {Teagle}(1999)}]{Alt:Wsea}
{Alt}, J.~C., \& {Teagle}, D.~A.~H. 1999, \gca, 63, 1527

\bibitem[{{Baraffe} {et~al.}(2015){Baraffe}, {Homeier}, {Allard}, \&
  {Chabrier}}]{Baraffe:2015}
{Baraffe}, I., {Homeier}, D., {Allard}, F., \& {Chabrier}, G. 2015, \aap, 577,
  A42

\bibitem[{{Barnes} {et~al.}(2009){Barnes}, {Jackson}, {Greenberg}, \&
  {Raymond}}]{Barnes:TidalLimits}
{Barnes}, R., {Jackson}, B., {Greenberg}, R., \& {Raymond}, S.~N. 2009, \apjl,
  700, L30

\bibitem[{{Barnes} {et~al.}(2013){Barnes}, {Mullins}, {Goldblatt}, {Meadows},
  {Kasting}, \& {Heller}}]{TidalVenuses}
{Barnes}, R., {Mullins}, K., {Goldblatt}, C., {et~al.} 2013, Astrobiology, 13,
  225

\bibitem[{{Berner}(2004)}]{Berner:Book}
{Berner}, R.~A. 2004, {The Phanerozoic Carbon Cycle}, 158

\bibitem[{{Berner} \& {Caldeira}(1997)}]{Berner:1997}
{Berner}, R.~A., \& {Caldeira}, K. 1997, Geology, 25, 955

\bibitem[{{Bolmont} {et~al.}(2017){Bolmont}, {Selsis}, {Owen}, {Ribas},
  {Raymond}, {Leconte}, \& {Gillon}}]{Bolmont:AO}
{Bolmont}, E., {Selsis}, F., {Owen}, J.~E., {et~al.} 2017, \mnras, 464, 3728

\bibitem[{{Brady} \& {G{\'{\i}}slason}(1997)}]{BradyGislason:1997}
{Brady}, P.~V., \& {G{\'{\i}}slason}, S.~R. 1997, \gca, 61, 965

\bibitem[{{Carr} {et~al.}(1998){Carr}, {McEwen}, {Howard}, {Chuang}, {Thomas},
  {Schuster}, {Oberst}, {Neukum}, {Schubert}, \& {Galileo Imaging
  Team}}]{Carr:Io}
{Carr}, M.~H., {McEwen}, A.~S., {Howard}, K.~A., {et~al.} 1998, \icarus, 135,
  146

\bibitem[{{Checlair} {et~al.}(2017){Checlair}, {Menou}, \&
  {Abbot}}]{Checlair:Mstar}
{Checlair}, J., {Menou}, K., \& {Abbot}, D.~S. 2017, \apj, 845, 132

\bibitem[{{Coogan} \& {Dosso}(2015)}]{CooganDosso:2015}
{Coogan}, L.~A., \& {Dosso}, S.~E. 2015, Earth and Planetary Science Letters,
  415, 38

\bibitem[{{Coogan} \& {Gillis}(2013)}]{CooganGillis:2013}
{Coogan}, L.~A., \& {Gillis}, K.~M. 2013, Geochemistry, Geophysics, Geosystems,
  14, 1771

\bibitem[{{Davies} \& {Davies}(2010)}]{q_earth}
{Davies}, J.~H., \& {Davies}, D.~R. 2010, Solid Earth, 1, 5

\bibitem[{{Driscoll} \& {Barnes}(2015)}]{Driscoll:2015}
{Driscoll}, P.~E., \& {Barnes}, R. 2015, Astrobiology, 15, 739

\bibitem[{{Dziewonski} \& {Anderson}(1981)}]{PREM}
{Dziewonski}, A.~M., \& {Anderson}, D.~L. 1981, Physics of the Earth and
  Planetary Interiors, 25, 297

\bibitem[{{Foley}(2015)}]{Foley:Area}
{Foley}, B.~J. 2015, \apj, 812, 36

\bibitem[{{Foley} {et~al.}(2012){Foley}, {Bercovici}, \& {Landuyt}}]{Foley:PT}
{Foley}, B.~J., {Bercovici}, D., \& {Landuyt}, W. 2012, Earth and Planetary
  Science Letters, 331, 281

\bibitem[{{Foley} \& {Smye}(2017)}]{Foley:2017}
{Foley}, B.~J., \& {Smye}, A.~J. 2017, ArXiv e-prints

\bibitem[{Gaillardet {et~al.}(1999)Gaillardet, Dupré, Louvat, \&
  Allègre}]{Gaillardet:weathering}
Gaillardet, J., Dupré, B., Louvat, P., \& Allègre, C. 1999, 159, 3

\bibitem[{{Gillon} {et~al.}(2017){Gillon}, {Triaud}, {Demory}, {Jehin}, {Agol},
  {Deck}, {Lederer}, {de Wit}, {Burdanov}, {Ingalls}, {Bolmont}, {Leconte},
  {Raymond}, {Selsis}, {Turbet}, {Barkaoui}, {Burgasser}, {Burleigh}, {Carey},
  {Chaushev}, {Copperwheat}, {Delrez}, {Fernandes}, {Holdsworth}, {Kotze}, {Van
  Grootel}, {Almleaky}, {Benkhaldoun}, {Magain}, \& {Queloz}}]{Trappist-1}
{Gillon}, M., {Triaud}, A.~H.~M.~J., {Demory}, B.-O., {et~al.} 2017, \nat, 542,
  456

\bibitem[{{Hansen} \& {Murray}(2012)}]{Hansen:2012}
{Hansen}, B.~M.~S., \& {Murray}, N. 2012, \apj, 751, 158

\bibitem[{{Hansen} \& {Murray}(2015)}]{Hansen:2015}
---. 2015, \mnras, 448, 1044

\bibitem[{{Haqq-Misra} {et~al.}(2016){Haqq-Misra}, {Kopparapu}, {Batalha},
  {Harman}, \& {Kasting}}]{Haqq-Misra:EBM}
{Haqq-Misra}, J., {Kopparapu}, R.~K., {Batalha}, N.~E., {Harman}, C.~E., \&
  {Kasting}, J.~F. 2016, \apj, 827, 120

\bibitem[{{Hoffman} \& {Schrag}(2002)}]{Hoffman:snowballEarth}
{Hoffman}, P.~F., \& {Schrag}, D.~P. 2002, Terra Nova, 14, 129

\bibitem[{{Howard} {et~al.}(2010){Howard}, {Marcy}, {Johnson}, {Fischer},
  {Wright}, {Isaacson}, {Valenti}, {Anderson}, {Lin}, \& {Ida}}]{Howard:2010}
{Howard}, A.~W., {Marcy}, G.~W., {Johnson}, J.~A., {et~al.} 2010, Science, 330,
  653

\bibitem[{{Jaeger} {et~al.}(2003){Jaeger}, {Turtle}, {Keszthelyi}, {Radebaugh},
  {McEwen}, \& {Pappalardo}}]{Jaeger:IoPlate}
{Jaeger}, W.~L., {Turtle}, E.~P., {Keszthelyi}, L.~P., {et~al.} 2003, Journal
  of Geophysical Research (Planets), 108, 12

\bibitem[{{Kasting} {et~al.}(1993){Kasting}, {Whitmire}, \&
  {Reynolds}}]{Kasting:1993}
{Kasting}, J.~F., {Whitmire}, D.~P., \& {Reynolds}, R.~T. 1993, \icarus, 101,
  108

\bibitem[{{Korenaga}(2010)}]{Korenaga:PT}
{Korenaga}, J. 2010, \apjl, 725, L43

\bibitem[{{Krissansen-Totton} \& {Catling}(2017)}]{KT_Catling:2017}
{Krissansen-Totton}, J., \& {Catling}, D.~C. 2017, Nature Communications, 8,
  15423

\bibitem[{{Kump} \& {Arthur}(1999)}]{KumpArthur}
{Kump}, L.~R., \& {Arthur}, M.~A. 1999, Chemical Geology, 161, 181

\bibitem[{{Maher} \& {Chamberlain}(2014)}]{MaherChamberlain}
{Maher}, K., \& {Chamberlain}, C.~P. 2014, Science, 343, 1502

\bibitem[{{McEwen} {et~al.}(2004{\natexlab{a}}){McEwen}, {Keszthelyi}, {Lopes},
  {Schenk}, \& {Spencer}}]{Io:Book}
{McEwen}, A.~S., {Keszthelyi}, L.~P., {Lopes}, R., {Schenk}, P.~M., \&
  {Spencer}, J.~R. 2004{\natexlab{a}}, {The lithosphere and surface of Io}, ed.
  F.~{Bagenal}, T.~E. {Dowling}, \& W.~B. {McKinnon}, 307--328

\bibitem[{{McEwen} {et~al.}(2004{\natexlab{b}}){McEwen}, {Keszthelyi}, {Lopes},
  {Schenk}, \& {Spencer}}]{Io:BookChapter}
---. 2004{\natexlab{b}}, {The lithosphere and surface of Io}, ed. F.~{Bagenal},
  T.~E. {Dowling}, \& W.~B. {McKinnon}, 307--328

\bibitem[{{Menou}(2015)}]{Menou:2015}
{Menou}, K. 2015, Earth and Planetary Science Letters, 429, 20

\bibitem[{{Mills} {et~al.}(2014){Mills}, {Lenton}, \& {Watson}}]{Mills:2014}
{Mills}, B., {Lenton}, T.~M., \& {Watson}, A.~J. 2014, Proceedings of the
  National Academy of Science, 111, 9073

\bibitem[{{Murray} \& {Dermott}(1999)}]{MurrayDermott}
{Murray}, C.~D., \& {Dermott}, S.~F. 1999, {Solar system dynamics}

\bibitem[{{Noack} \& {Breuer}(2014)}]{Noack:PT}
{Noack}, L., \& {Breuer}, D. 2014, \planss, 98, 41

\bibitem[{{Noack} {et~al.}(2017){Noack}, {Rivoldini}, \& {Van
  Hoolst}}]{Noack:melt}
{Noack}, L., {Rivoldini}, A., \& {Van Hoolst}, T. 2017, Physics of the Earth
  and Planetary Interiors, 269, 40

\bibitem[{{Ohtani} {et~al.}(1995){Ohtani}, {Nagata}, {Suzuki}, \&
  {Kato}}]{Ohtani:melt}
{Ohtani}, E., {Nagata}, Y., {Suzuki}, A., \& {Kato}, T. 1995, Chemical Geology,
  120, 207

\bibitem[{{O'Neill} {et~al.}(2007){O'Neill}, {Jellinek}, \&
  {Lenardic}}]{ONeill:PT}
{O'Neill}, C., {Jellinek}, A.~M., \& {Lenardic}, A. 2007, Earth and Planetary
  Science Letters, 261, 20

\bibitem[{{O'Reilly} \& {Davies}(1981)}]{OD:Io}
{O'Reilly}, T.~C., \& {Davies}, G.~F. 1981, \grl, 8, 313

\bibitem[{{Owen} {et~al.}(1979){Owen}, {Cess}, \& {Ramanathan}}]{Owen:1979}
{Owen}, T., {Cess}, R.~D., \& {Ramanathan}, V. 1979, \nat, 277, 640

\bibitem[{{Phillips} {et~al.}(2000){Phillips}, {McEwen}, {Keszthelyi},
  {Geissler}, {Simonelli}, {Milazzo}, \& {Galileo Imaging
  Team}}]{Io:resurfacing}
{Phillips}, C.~B., {McEwen}, A.~S., {Keszthelyi}, L.~P., {et~al.} 2000, in
  Bulletin of the American Astronomical Society, Vol.~32, AAS/Division for
  Planetary Sciences Meeting Abstracts \#32, 1046

\bibitem[{{Rye} {et~al.}(1995){Rye}, {Kuo}, \& {Holland}}]{Rye:1995}
{Rye}, R., {Kuo}, P.~H., \& {Holland}, H.~D. 1995, \nat, 378, 603

\bibitem[{{Sagan} \& {Mullen}(1972{\natexlab{a}})}]{Sagan:1972}
{Sagan}, C., \& {Mullen}, G. 1972{\natexlab{a}}, Science, 177, 52

\bibitem[{{Sagan} \& {Mullen}(1972{\natexlab{b}})}]{Sagan:NH3}
---. 1972{\natexlab{b}}, Science, 177, 52

\bibitem[{{Sakamaki} {et~al.}(2006){Sakamaki}, {Suzuki}, \&
  {Ohtani}}]{Sakamaki:melt}
{Sakamaki}, T., {Suzuki}, A., \& {Ohtani}, E. 2006, \nat, 439, 192

\bibitem[{{Schaefer} {et~al.}(2016){Schaefer}, {Wordsworth}, {Berta-Thompson},
  \& {Sasselov}}]{Schaefer:AO}
{Schaefer}, L., {Wordsworth}, R.~D., {Berta-Thompson}, Z., \& {Sasselov}, D.
  2016, \apj, 829, 63

\bibitem[{{Sleep} \& {Zahnle}(2001)}]{Sleep:2001}
{Sleep}, N.~H., \& {Zahnle}, K. 2001, \jgr, 106, 1373

\bibitem[{{Spencer} {et~al.}(2000){Spencer}, {Rathbun}, {Travis}, {Tamppari},
  {Barnard}, {Martin}, \& {McEwen}}]{Io:Spencer2000}
{Spencer}, J.~R., {Rathbun}, J.~A., {Travis}, L.~D., {et~al.} 2000, Science,
  288, 1198

\bibitem[{{Stacey} \& {Davis}(2008)}]{PhysicsEarth}
{Stacey}, F.~D., \& {Davis}, P.~M. 2008, {Physics of the Earth}

\bibitem[{{Stamenkovi{\'c}} {et~al.}(2012){Stamenkovi{\'c}}, {Noack}, {Breuer},
  \& {Spohn}}]{Stamen:PT}
{Stamenkovi{\'c}}, V., {Noack}, L., {Breuer}, D., \& {Spohn}, T. 2012, \apj,
  748, 41

\bibitem[{{Tackley} {et~al.}(2013){Tackley}, {Ammann}, {Brodholt}, {Dobson}, \&
  {Valencia}}]{Tackley:PT}
{Tackley}, P.~J., {Ammann}, M., {Brodholt}, J.~P., {Dobson}, D.~P., \&
  {Valencia}, D. 2013, \icarus, 225, 50

\bibitem[{{Tamayo} {et~al.}(2017){Tamayo}, {Rein}, {Petrovich}, \&
  {Murray}}]{Tamayo:2017}
{Tamayo}, D., {Rein}, H., {Petrovich}, C., \& {Murray}, N. 2017, \apjl, 840,
  L19

\bibitem[{{Tian}(2015)}]{Tian:AO}
{Tian}, F. 2015, Earth and Planetary Science Letters, 432, 126

\bibitem[{{Turcotte} \& {Schubert}(2002)}]{Geodynamics:Book}
{Turcotte}, D.~L., \& {Schubert}, G. 2002, {Geodynamics - 2nd Edition}, 472

\bibitem[{{Urey}(1952)}]{Urey:CO2}
{Urey}, H.~C. 1952, Proceedings of the National Academy of Science, 38, 351

\bibitem[{{Valencia} {et~al.}(2006){Valencia}, {O'Connell}, \&
  {Sasselov}}]{Valencia:2006}
{Valencia}, D., {O'Connell}, R.~J., \& {Sasselov}, D. 2006, \icarus, 181, 545

\bibitem[{{Valencia} {et~al.}(2007){Valencia}, {O'Connell}, \&
  {Sasselov}}]{Valencia:PT}
{Valencia}, D., {O'Connell}, R.~J., \& {Sasselov}, D.~D. 2007, \apjl, 670, L45

\bibitem[{{van Heck} \& {Tackley}(2011)}]{VanHeck:PT}
{van Heck}, H.~J., \& {Tackley}, P.~J. 2011, Earth and Planetary Science
  Letters, 310, 252

\bibitem[{{Veeder} {et~al.}(1994){Veeder}, {Matson}, {Johnson}, {Blaney}, \&
  {Goguen}}]{q_Io:Veeder}
{Veeder}, G.~J., {Matson}, D.~L., {Johnson}, T.~V., {Blaney}, D.~L., \&
  {Goguen}, J.~D. 1994, \jgr, 99, 17095

\bibitem[{{Vinson} \& {Hansen}(2017)}]{Hansen:2017}
{Vinson}, A.~M., \& {Hansen}, B.~M.~S. 2017, ArXiv e-prints

\bibitem[{{Walker} {et~al.}(1981){Walker}, {Hays}, \& {Kasting}}]{Walker:1981}
{Walker}, J.~C.~G., {Hays}, P.~B., \& {Kasting}, J.~F. 1981, \jgr, 86, 9776

\bibitem[{{West}(2012)}]{West:2012}
{West}, A.~J. 2012, Geology, 40, 811

\bibitem[{{West} {et~al.}(2005){West}, {Galy}, \& {Bickle}}]{West:2005}
{West}, A.~J., {Galy}, A., \& {Bickle}, M. 2005, Earth and Planetary Science
  Letters, 235, 211

\bibitem[{{Williams} \& {Kasting}(1997)}]{WilliamKasting:1997}
{Williams}, D.~M., \& {Kasting}, J.~F. 1997, \icarus, 129, 254

\end{thebibliography}

\end{document}